\documentclass[mathleft
]{an}
\usepackage{graphicx}
\usepackage{times}
\usepackage{natbib,amsmath}
\def\celsius{^\circ\rm C}
\def\deg{$^\circ$}
\def\wig#1{\mathrel{\hbox{\hbox to 0pt{%
          \lower.6ex\hbox{$\sim$}\hss}\raise.4ex\hbox{$#1$}}}}
\def\focpos{p_{\rm FocPos}}
\def\fwhm{{\cal W}}
\def\meanrhosq{{\langle \rho'^2 \rangle}}
\def\turbseeing{{\cal S}_{\rm turb}}
\def\alignl#1{\multicolumn{1}{l}{#1}}
\begin{document}

\Pagespan{1}{}
\Yearpublication{2015}%
\Yearsubmission{2015}%
\Month{2}%
\Volume{}%
\Issue{}%

\title{Thermalizing a telescope in Antarctica \\ Analysis of ASTEP observations}

\author{T. Guillot\inst{1}\fnmsep\thanks{Corresponding author:
  \email{tristan.guillot@oca.eu}\newline}
L. Abe\inst{1}, A. Agabi\inst{1}, J.-P. Rivet\inst{1}, J.-B. Daban\inst{1}, D. M\'ekarnia\inst{1}, E. Aristidi\inst{1}, F.-X. Schmider\inst{1}, N. Crouzet\inst{2}, I. Gon\c{c}alves\inst{1}, C. Gouvret\inst{1}, S. Ottogalli\inst{1}, H. Faradji\inst{1},  P.-E. Blanc\inst{3}, E. Bondoux\inst{1}, F. Valbousquet\inst{4}
}
\titlerunning{Thermalizing a telescope in Antarctica}
\authorrunning{T. Guillot et al. }
\institute{
\inst{1}Laboratoire Lagrange, Universit\'e de Nice-Sophia Antipolis, Observatoire de la C\^ote d'Azur, CNRS UMR 7293, Nice Cedex 4, France\\
\inst{2}Dunlap Institute for Astronomy \& Astrophysics, University of Toronto, 50 St. George Street, Toronto, Ontario M5S 3H4, Canada\\
\inst{3}Observatoire de Haute-Provence, Universit\'e d'Aix-Marseille \& CNRS, 04870 Saint Michel l'Observatoire, France\\
\inst{4}Optique et Vision, 6 bis avenue de l'Est\'erel, BP 69, 06162 Juan-Les-Pins, France}

\received{February 11, 2015, revised June 11, 2015}
\accepted{June 15, 2015}
\publonline{}

\keywords{instrumentation: photometers, techniques: photometric, telescopes, atmospheric effects}

\abstract{%
  The installation and operation of a telescope in Antarctica represent particular challenges, in particular the requirement to operate at extremely cold temperatures, to cope with rapid temperature fluctuations and to prevent frosting. Heating of electronic subsystems is a necessity, but solutions must be found to avoid the turbulence induced by temperature fluctuations on the optical paths. ASTEP~400 is a 40 cm Newton telescope installed at the Concordia station, Dome C since 2010 for photometric observations of fields of stars and their exoplanets. While the telescope is designed to spread star light on several pixels to maximize photometric stability, we show that it is nonetheless sensitive to the extreme variations of the seeing at the ground level (between about 0.1 and 5 arcsec) and to temperature fluctuations between $-30^\circ$C and $-80^\circ$C. We analyze both day-time and night-time observations and obtain the magnitude of the seeing caused by the mirrors, dome and camera. The most important effect arises from the heating of the primary mirror which gives rise to a mirror seeing of $0.23\,\rm arcsec\,K^{-1}$. We propose solutions to mitigate these effects.}

\maketitle

\section{Introduction}

Operating a telescope in Antarctica and in particular on one of its high altitude plateaus such as Dome C is a formidable opportunity due to the continuous winter night, excellent weather \citep{Crouzet+2010}, low water abundance \citep{Burton2010}, low scintillation \citep{Kenyon+2006}, and also with the perspective of extremely low turbulence once above a $\sim 30$\,meters-thick boundary layer \citep{Trinquet+2008,Aristidi+2009}. 
It is also a great challenge because of the remoteness of the continent, low temperatures (down to -80$^\circ$C in the winter), presence of ice and frosting of the instruments and the generally limited internet connexion which requires on-site, fast treatment of the data. 

ASTEP~400 is a 40cm telescope installed since 2010 at the Concordia station located at -75.06$^\circ$S, 123.3$^\circ$E and an altitude of 3233 meters. ASTEP (Antarctica Search for Transiting ExoPlanets) is a pilot project to both characterize the Dome C site for photometric surveys and discover and characterize transiting exoplanets through accurate, visible photometry \citep{Fressin+2005}. The detection of the secondary eclipse of planet WASP-19b {\em behind} its star (a mere 370\,ppm signal) \citep{Abe+2013}, a first time at these wavelengths with a ground-based telescope is a testimony to the high photometric quality of the site and the possibility to perform excellent observations there. However, no new transiting planets have been detected so far, in spite of an excellent overall duty cycle during these four years of operation and tens of candidates, some of which are still being followed up (M\'ekarnia et al., in preparation). The reason for this is in part the delay in acquiring the data during the first seasons of ASTEP observations, the complex data pipeline which had to be set up, and as we will see the large point spread functions (PSFs) which mean a higher confusion with other stars, especially in crowded fields \citep[e.g.][]{Bachelet+2012}. A comparison of ASTEP and BEST II (Chile) shows a photometric quality that is superior for ASTEP for bright stars but lower for faint stars \citep{Fruth+2014}. The latter can probably be attributed to the large PSFs of ASTEP, which means that more photons are lost to the background in the case of faint stars. 

As we will see, these large PSFs are mostly due to the high level of turbulence of the boundary layer for a telescope which is installed only about 2 meters above the ground. Installing the telescope higher could thus be extremely beneficial. However, this requires understanding how temperature gradients and fluctuations also affect the results. This is also true for any optical or near-infrared telescope installed or to be installed on such a harsh environment \citep[e.g.][]{Tosti+2006,Strassmeier+2007,Strassmeier+2008,Chadid+2010,Zhou+2010,Burton+2010,Yuan+Su2012, Shang+2012,Chadid+2014}. This analysis can therefore be beneficial to other projects. 

We present in Section~2 the design of the ASTEP~400 telescope. We then present theoretical calculations of thermal deformations of the telescope and the expected consequences for turbulence on the optical path. In Section~4, we present specific tests conducted during the summer season to identify the subsystems influencing the image quality the most.  We then analyze globally the observations conducted during the first (winter) seasons of the instrument and combined to direct characterization of the atmospheres with a DIMM telescope. 

\section{ASTEP~400: design}

The optical and mechanical design of ASTEP~400 is described by \cite{Daban+2010}. Hereafter, we focus on aspects directly related to the consequences of mechanical and thermal changes of the structure of the instrument. 

\subsection{Telescope}

\subsubsection{Structural analysis overview}

The study of the ASTEP telescope concept was conducted with the purpose to minimize the photometric variations during the observations. It included in particular the effects of thermal variations and the flexures of the mechanical structure due to gravity in different positions. Thermo-mechanical studies were achieved using analytical calculation and PATRAN / NASTRAN finite element modelling.

We now present the detailed models of the structure of the telescope done with the finite element software. Although the structure of the instrument had to be simplified for this modeling, all the main elements were included for a precise estimate of the global behavior. 

\begin{table}
\caption{Material properties}\label{tab:material}
\begin{tabular}{llll}
\hline\hline
Material & Elasticity & Density & CTE \\
 & module & & \\
 & [MPa] & [kg/l] & [$10^{-6}\,\rm K^{-1}$] \\ \hline
TA6V (titanium alloy) & 110,000 & 4.5 & 8 \\
2017A (dural) & 73,000 & 2.8 & 22 \\
Epoxy carbon fiber & 125,000 & 1.55 & 0.25 \\
Zerodur & 90,000 & 2.53 & -0.1 \\
Invar & 145,000 & 8.0 & 2\\
Inox steel & 200 000 & 7.9 & 15\\
\hline\hline
\end{tabular}
\end{table}

Table~\ref{tab:material} provides a summary of the main properties of materials used in the telescope. Except for zerodur for which the coefficient of thermal expansion (CTE) corresponds to its value at a temperature of $-75^\circ$C, all the other properties refer to room temperature ($+20^\circ$C). The elasticity module $E$ tends to increase with decreasing temperature but above 200\,K this is relatively small. A test in a cold chamber with carbon tubes confirmed that to within $10\%$, $E$ did not vary between $+20^\circ$C and $-80^\circ$C. Variations in the CTE with temperature are not well-known but they are generally relatively small and may thus be neglected. 

Several cases were considered, with a choice between different structural elements, and different test cases for the telescope, i.e. with different positions during the night and at different temperatures. The calculated deformations, like in this example, were introduced as an input in the Zemax optical study, to see the effects of the thermal variations and flexures on the images along the night.

This study led to the choice of a Serrurier truss with carbon-epoxy tubes for the structure of the telescope, giving the best results with minimal weight \citep[see][for a description of the elements of the telescope]{Daban+2010}. It concluded that the main effects of the deformations were a translation of the field on the detector corresponding to 15 arcsec on the sky at maximum, displacement that is corrected by the fine guiding, and a maximum defocus of 50 µm, that could be also compensated by a change of the focus. Second order effects include deformations of the PSF and photometric variations.
Assuming an initial PSF width of 3 arcsec, the increase of PSF width should not exceed 2 - 3\% in the worst case. The associated photometric variations, calculated by the change in encircled energy, remain below 0.1\% inside an aperture of 6 pixels \citep{Daban+2010}.

\subsubsection{Structure modeling \& thermal dilatations}

\begin{figure}
\includegraphics[width=\hsize]{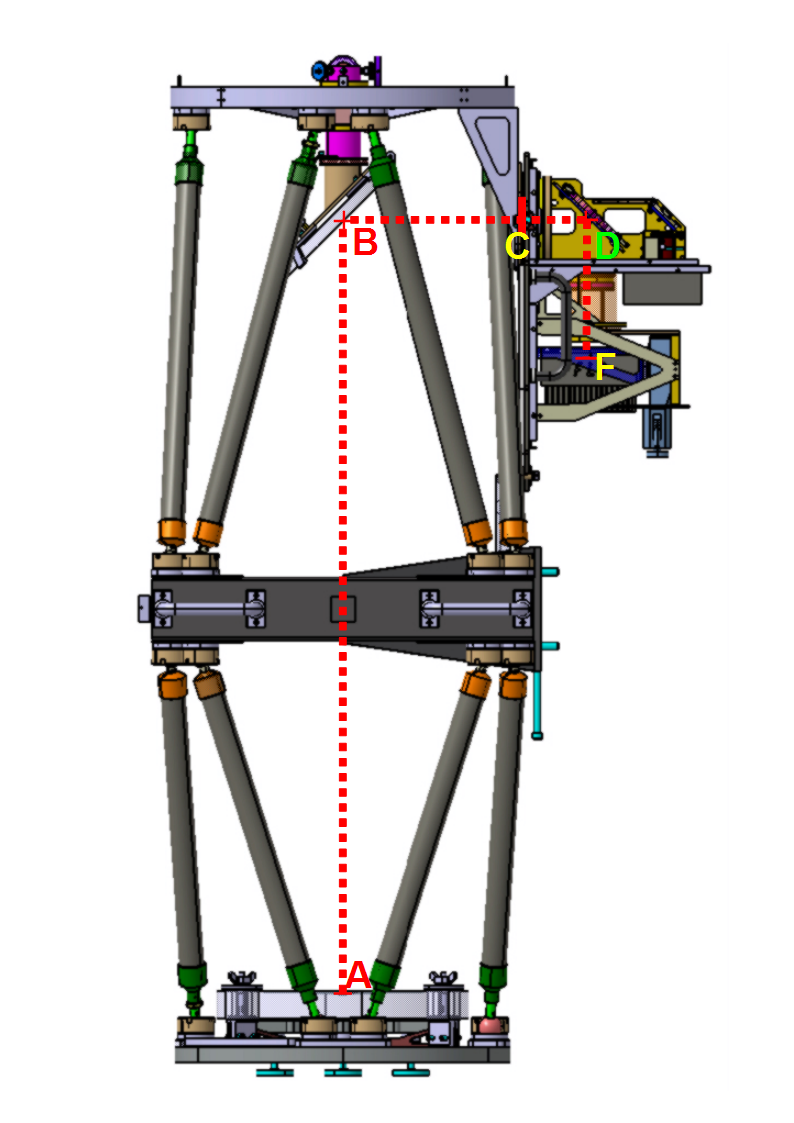}
\caption{Cut plane of the ASTEP~400 telescope and camera. The optical axis is defined by the points A, B, C, D and F. A is the optical center of the primary parabolic mirror. B is the optical center of the secondary mirror. C is the center of the entrance window of the camera box. D is the optical center of the dichroic mirror. And F is the center of the focal plane. The distances between these points are as follows: $\rm AB=1245\,mm$, $\rm BC=290\,mm$, $\rm CD=107\,mm$, $\rm DF=225\,mm$.}
\label{fig:ASTEP_dilatation1}
\end{figure}

Figure~\ref{fig:ASTEP_dilatation1} shows the mechanical concept of the optical tube assembly. It has been designed in order to minimize as far as possible the global coefficient of thermal expansion (CTE) between point A and point F. Analytical calculations and finite element analyses using NASTRAN have been done to estimate this global CTE. These calculations do not take into account the expansion of the window's lenses that could affect their optical power and then affect the focal plane position. We will see in Section~\ref{sec:thermoelastic} that this effect is small, but not completely negligible. 

Our analytical linear estimate using the CTE from table~\ref{tab:material} and the telescope structure \citep{Daban+2010} show that, for 30K temperature increase, the distance AB is reduced by $43\,\mu$m.
Indeed, the very low CTE carbon/epoxy bars of the Serrurier structure, associated with Aluminum alloy used in the primary mirror barrel, in the central frame, in the upper ring and in the secondary mirror support, lead to a negative global CTE for the segment [AB].
Since the upper ring of the telescope and the interface with camera box are aluminum parts, segment [BC] expends positively according to the aluminum CTE. Thus, for 30K temperature increase, distance BC expands by $+193\,\mu$m.
Then, inside the camera box, temperature changes are regulated and may not exceed $\pm5\,$K. Given that the mechanical structure holding the dichroic mirror D is built with titanium alloy, a temperature rise of $5\,$K will induce an extension of distance CD by 5$\,\mu$m.
Finally, the distance change between the dichroic mirror D and the focal plane F depends on the extension of the carbon fiber /epoxy plate that hold all the camera box components. This plate feels the $30\,$K outer temperature increase and DF is thus increased by 2$\,\mu$m.
As a conclusion, for a $30\,$K temperature increase, the expansion between A and F will be $-43 + 193 + 5 + 2 = 157\,\mu$m. We therefore estimate that a temperature variation yields a change of the focal plane by $\sim 5\rm\,\mu m\,K^{-1}$.

\begin{figure}
\includegraphics[width=\hsize]{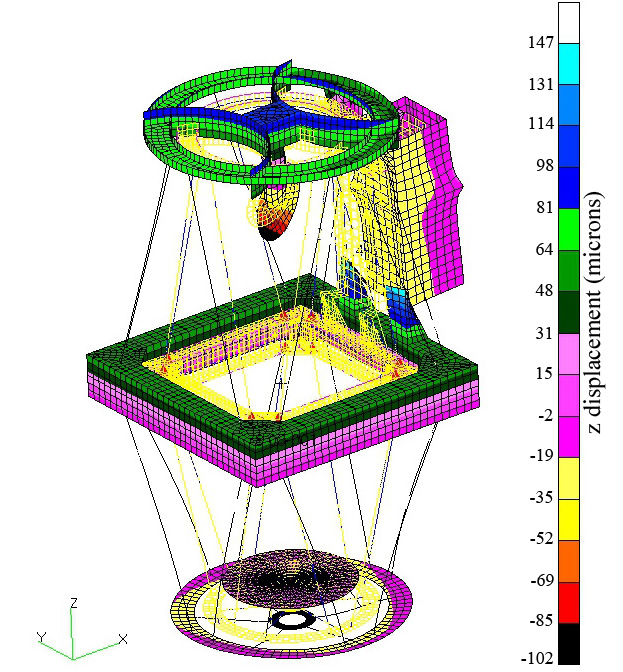}
\caption{Mechanical deformation of the telescope structure under the effect of the thermal dilatations of $+30\,$K relative to the nominal value. The colors show the vertical displacement ($z$ direction), which range between $-102$ and $+147\,\mu$m.}
\label{fig:thermo-elastic_iso}
\end{figure}

Figure~\ref{fig:thermo-elastic_iso} shows the resulting displacement of the telescope structure to a $+30\,$K temperature increase, using our NASTRAN simulations. As boundary condition, we fixed the central node of the M1 mirror both in translation and in rotation. The resulting variations agree with the analytical approach with changes in the focal plane of order $\sim 7\rm\,\mu m\,K^{-1}$. 

\subsubsection{Mechanical flexions: Structure modeling}

\begin{figure}
\includegraphics[width=\hsize]{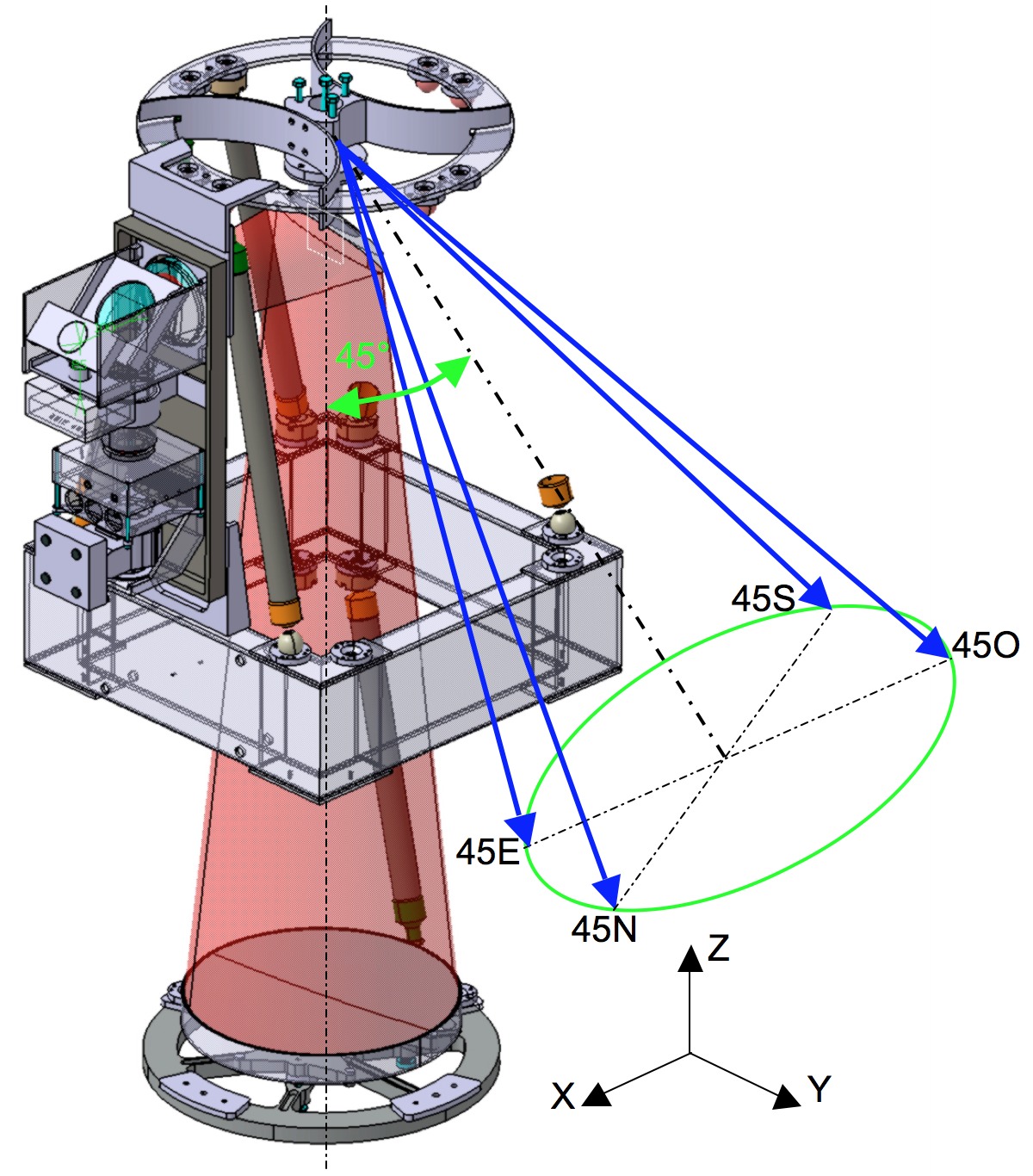}
\caption{Cutaway view of the ASTEP~400 telescope and orientation of the gravity vector for the four cases considered in this study.}
\label{fig:static_experiments}
\end{figure}

We now consider the deformations of the telescope resulting from its weight for four different pointings indicated in fig.~\ref{fig:static_experiments}. (The structure of the telescope used in this section results from a design that is slightly different than the final one, in particular with a thicker central case, but this is expected to have a negligible effect on the results.) These pointings account for the installation of the telescope at the Concordia station ($-75^\circ$ latitude) for a declination of $45^\circ$ towards the North, East, South and West, respectively. 

For the NASTRAN simulations, the boundary condition is that the face of the central housing tied to the mount is held fixed. 
For each pointing, the largest displacement is obtained at the outermost edge of the camera box. However, this displacement remains relatively similar for the four pointings considered so that the effect on the observations is limited.  All in all, the telescope weighs 83\,kg, including 23\,kg for the camera box.
%
%
Table~\ref{tab:differential} shows the resulting displacements and tilts relative to the 45N case at three specific locations on the optical path: at the center of the M1 mirror, at the center of the M2 mirror and at the camera box entrance. The change in tilt is small and is corrected by the telescope guiding. The displacements are limited to less than $40\,\mu$m in absolute positions. When one considers variations in the distance between the M1 and the camera box entrance, these are even smaller, i.e. $11\,\mu$m in the worst case ($\Delta X$, for the 45E pointing). Unlike thermal dilatations of the instrument, its flexions may be neglected for our purposes. 

\begin{table*}
\caption{Differential displacements}\label{tab:differential}
\begin{tabular}{llrrrrrr}
\hline\hline
\alignl{Pointing} &\alignl{Element}& \alignl{$\Delta X\rm\ [\mu m]$} & \alignl{$\Delta Y\rm\ [\mu m]$} & \alignl{$\Delta Z\rm\ [\mu m]$} & \alignl{Tilt$_X\ [^\circ]$} & \alignl{Tilt$_Y\ [^\circ]$} & \alignl{Tilt$_Z\ [^\circ]$} \\ \hline
45W & M1 & $-20$ & $12$ & $4$ & $0.000$ & $0.000$ & $0.000$  \\
       & M2 &  $-35$ & $9$ & $4$ & $0.001$ & $0.003$ & $0.001$  \\
       & CamBox & $-20$ & $11$ & $3$ & $-0.001$ & $-0.002$ & $0.002$ \\ \hline
45S & M1 & $-4$ & $33$ & $7$ & $-0.001$ & $0.000$ & $-0.001$  \\
       & M2 &  $-3$ & $34$ & $7$ & $0.001$ & $0.000$ & $0.000$  \\
       & CamBox & $0$ & $28$ & $6$ & $-0.003$ & $0.000$ & $0.004$ \\ \hline
45E & M1 & $12$ & $11$ & $4$ & $0.000$ & $0.001$ & $0.000$  \\
       & M2 &  $36$ & $9$ & $4$ & $0.000$ & $-0.002$ & $0.001$  \\
       & CamBox & $23$ & $11$ & $4$ & $-0.001$ & $0.002$ & $0.002$ \\ \hline
\hline
\end{tabular}
\end{table*}


\subsection{Camera box}

The camera box was designed to minimize temperature fluctuations and turbulence on the optical path while maintaining some subsystems above $0^\circ$C with a much colder outside temperature.

\begin{figure}
\includegraphics[width=\hsize]{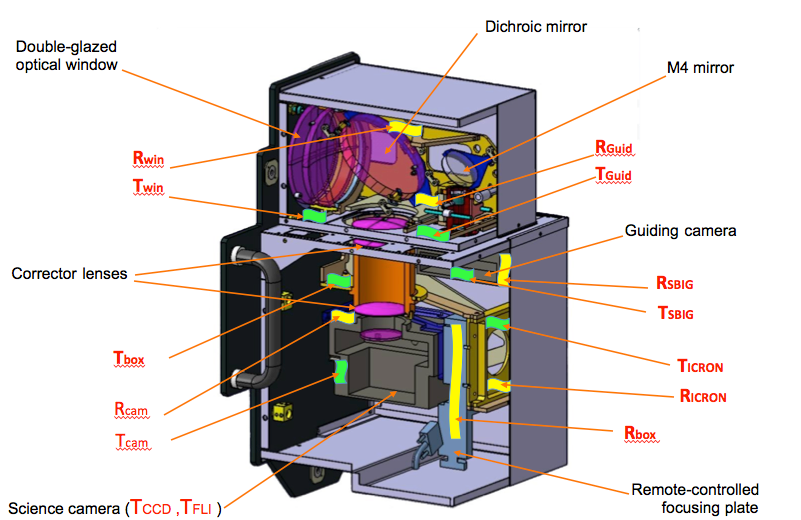}
\caption{Cutaway view of the ASTEP~400 camera box with the main optical and electronic parts. The resistances are labeled (``R'') and the thermal probes (``T'').}
\label{fig:focal-box}
\end{figure}

In order to maintain optimal functioning temperature for each electronic part (science and guiding cameras, microcontrol translation platform, conversion modules for the camera data), the thermal insulation of the camera box is ensured by individual heating modules. Figure~\ref{fig:focal-box} shows the positions of the heating components on the different parts of the camera box. Each component is formed by a resistance and a temperature probe. The power of the resistance is determined as a function of the goal temperature by a PID (proportional-integral-derivative) controller.

\section{Optical consequences of thermoelastic distortions}\label{sec:thermoelastic}

\subsection{Mirror defrosting}

The defrosting device for the primary and secondary mirrors (M1 and M2) of ASTEP~400 involves
custom-designed planar heaters (Inconel600\textsuperscript{\textregistered} stripes
sandwiched between Kapton\textsuperscript{\textregistered} sheets)
in thermal contact with the rear faces of the mirrors. These heating resistors are powered
by software-driven Pulse Width Modulation (PWM) controllers delivering a tunable
average power between $0\%$ and $100\%$ of a maximum value ($250$\,W for M1, and $115$\,W for M2).

Small heating power ratios (below $10\%$) are used continuously,
to avoid frost deposit (``preventive mode''). However, when the external temperature rises too rapidly,
the mirror's thermal inertia may lead its optical surface to be temporarily cooler than the frost
point. In this situation, frost deposit can occur even if the heaters are
in preventive mode. To remove it without on-site mechanical action,
the heaters can be set to a ``curative mode'' (power ratios between $50\%$ and $100\%$)
for a short time (less than one hour). Of course, image acquisition must be stopped,
because a strong heating would severely hamper the image quality.

In the ``normal'' (preventive) mode, the power fraction delivered to the heaters must
be carefully tuned to prevent frost formation without damaging too much the image
quality. This is done empirically.

The image quality degradation induced by the defrosting heaters can have three distinct origins:
{(i)} turbulence production within the instrument, {(ii)} thermoelastic
distortion of telescope's structure, {(iii)} thermoelastic
distortion of the mirrors themselves. As far as the telescope's structure is concerned, special
care have been taken by the designers to reduce its effects on the images quality, and only a
slight focus shift is likely to happen. Thus, a motorized stage have been introduced
to compensate for it.

To reduce the effects of thermoelastic distortion of the mirrors themselves, a low
expansion coefficient material, the Schott Zerodur\textsuperscript{\textregistered}
(grade 2) has been chosen. This material has very low (and even slightly negative)
thermal expansion coefficients in a wide range of temperatures.
However, since the mirrors have been polished and optically controlled at usual
``room temperature'' (some $20^\circ$C), but used at temperatures as low as $-70^\circ$C
or below, the cumulative effect of thermoelastic distortion on the optical surface
needed to be investigated, to figure out its relative importance on the PSF widening.
Since the various sources of heater-induced PSF degradations (convective turbulence,
telescope's structure distortion, and mirror distortion) are quite difficult to disentangle
one from the others experimentally, we have addressed this issue by
finite element numerical simulations coupled to ray-tracing computations. Our goal was not
to reach state-of-the-art accurate values, but rather to get some insight on the
orders of magnitude. Our computations incorporate the following effects:
\begin{itemize}\itemsep=0pt
\item
Radiative cooling. Ambient temperature is chosen to be $200$\,K, a value frequently reached
at Dome~C during winter). The protected aluminium coated reflective surface is assumed
to have an emissivity coefficient of $0.12$ ($88\%$ reflectivity in the visible domain\,;
no transmission), which is a commonly used value. For the side
and rear surfaces however, the emissivity is not known accurately. Thus, we chose the
``{\sl bona fide}'' value $0.50$.
\item
Natural convection. Obtaining an accurate value for the thermal convection coefficient
in not a trivial issue \citep{W07a}. Since the convection is natural (no fan in the telescope's tube),
this coefficient is known to range from $5\rm\,W\,m^{-2}\,K^{-1}$ to $15\rm\,W\,m^{-2}\,K^{-1}$.
For a ``worst case'' simulation maximizing thermoelastic effects, we have chosen
the lower value: $h=5\rm\,W\,m^{-2}\,K^{-1}$.
\item
Uniform surface heat flux on the bottom face from the defrosting heater.
\item
The mirrors are supposed to be optically perfect (or at least diffraction-limited)
at a reference temperature of $293$\,K, with is a reasonable estimate of the
temperature at which they were controlled by the manufacturer.
\item
The regime is assumed to be stationary (external temperature and heater power
ratio assumed to be constant).
\item
To account for the thermal relative length variations of Zerodur\textsuperscript{\textregistered}
between $293$\,K (manufacturing temperature) and $200$\,K (operating temperature),
the average value of $-1.\,10^{-7}$\,K$^{-1}$ have been retained for the linear
thermal expansion coefficient.
\end{itemize}

\subsection{The primary mirror}\label{sec:M1}

The primary mirror of ASTEP~400 is parabolic, with a radius of curvature of $3730$\,mm, a mechanical diameter of $405$\,mm, and an edge thickness of $45$\,mm. Finite elements simulations of thermoelastic deformations resulting from the resistor heating were done for power ratios
$0\%$, $5\%$, $10\%$, $15\%$, $20\%$, $30\%$, $40\%$, $50\%$, $60\%$, $70\%$, $80\%$, $90\%$,
and $100\%$ of the $250$\,W maximal power. The finite elements geometry in the $(r,z)$ half-plane (assuming axial symmetry)
involves $1009$ nodes and $1832$ elements, with an average element distortion index of $0.85$.

\begin{figure}
   \includegraphics[width=\hsize]{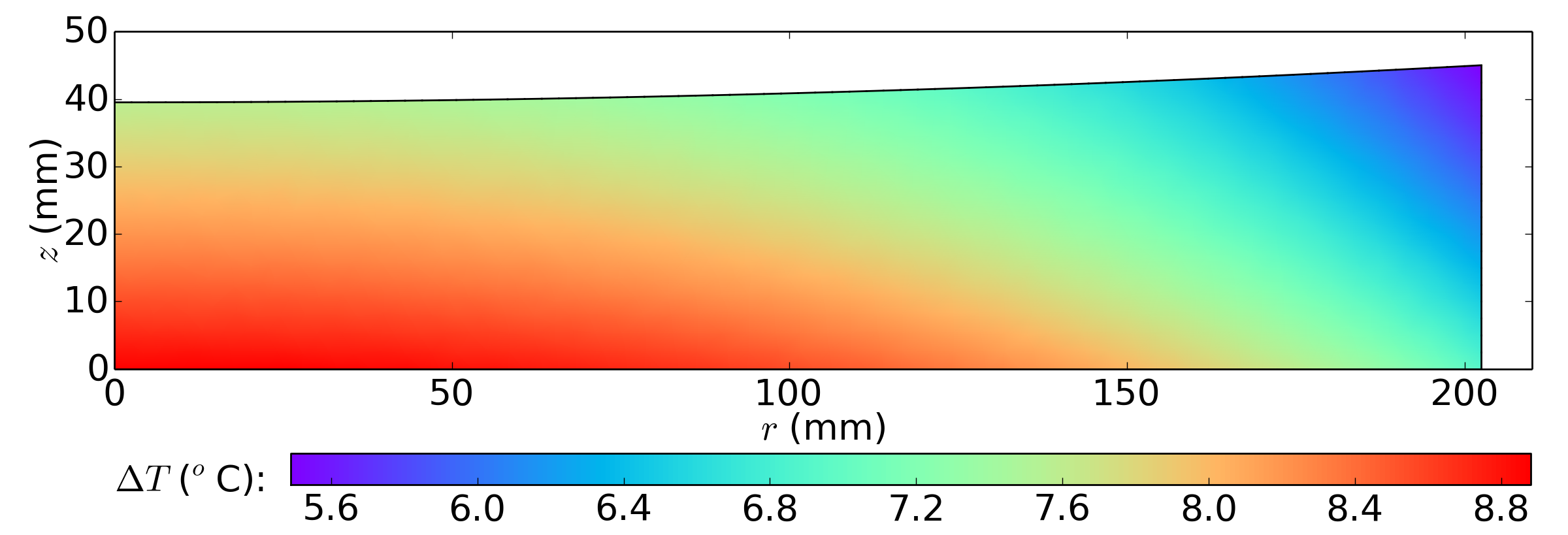}
   \includegraphics[width=\hsize]{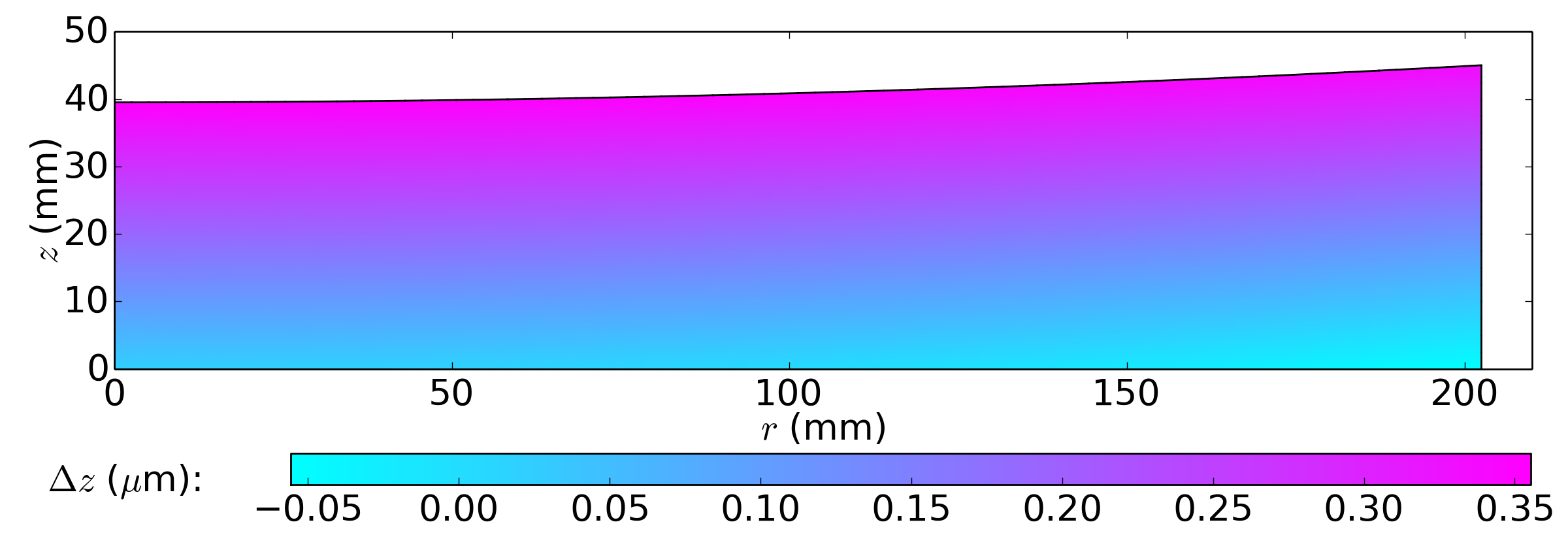}
   \caption{
Thermoelastic effects on M1 for a $10\%$ defrosting power. The pictures are meridian cuts through the primary mirror,
and show half of it only, because of its axis-symmetry.
   {\em Upper panel\/}: Temperature deviation (from 5.5$^\circ$C to 8.9$^\circ$C). {\em Lower panel\/}: vertical distortion (from $-0.06\,\mu$m to $0.36\,\mu$m). The parameters used for the calculations are $T_0=293\,$K, $T_{\rm ext}=200\,$K, $\epsilon_0=0.12$, $\epsilon_1=0.50$, $h=5.0\rm\,W\,m^{-2}\,K^{-1}$, and $P=10\%$.}
   \label{fig:M1_TemperatureDisplacement}
\end{figure}

As a sample, the temperature field and $z$-displacement fields for a power ratio
of $10\%$ are shown in Figure~\ref{fig:M1_TemperatureDisplacement}.
Such a power ratio is a typical value of the power ratio used at
Dome~C in curative mode.
For the temperature field, the external temperature ($200$\,K) has been subtracted.
The reference state for the node's $z$-displacement is the mirror at manufacturing
temperature ($293$\,K). With this power rate, the simulated temperature elevations are
within $+5.6^\circ$C  and  $8.8^\circ$C above ambiant. These values are in qualitative agreement with the
data measured at Dome~C instable conditions, by a PT100 probe glued on the side
of the cylindrical surface of the primary mirror. According to this simulation, the values of the
$z$-displacement in the primary mirror range from $-0.05$\,$\mu$m to  $+0.35$\,$\mu$m,
which is {\sl a priori} non negligible for an optical surface operated in the visible domain.

To get a better insight on the optical consequences of such displacements, we have performed
a ray-tracing computation to estimate their effect on the on-axis PSF size. This reveals that
the main effect of thermoelastic distortion is a focus shift smaller than $0.2$\,mm for
power ratios between $0\%$ and $90\%$. This can be easily compensated for by displacing
the sensor (mounted on a motorized stage), without hampering noticeably the efficiency
of the 5-lenses coma corrector. In addition to this slight defocus, thermoelastic distortions
produce residual aberrations (less than $0.1''$ for all power ratios), which remain
below the diffraction limit ($0.44''$ at $700$\,nm).

\subsection{The secondary mirror}
The problem of the secondary mirror is slightly different. As in standard Newton telescopes,
it is flat
(semi-major axis: $236$\,mm; semi-minor axis: $166$\,mm; thickness: $20$\,mm).
It is inclined at $45^\circ$ with respect to the telescope's main optical axis. Thus, even
a slight thermally-induced curvature is likely to introduce both defocus and astigmatism.

The same kind of analysis has been performed for the secondary mirror. This reveals that
a slight focus shift is introduced by the thermoelastic distortion (less than $0.05$~mm for
all power ratios). As for the primary mirror, the focus shift remains in the tolerance range
of the coma corrector, and can be compensated for by an appropriate sensor shift. However,
besides the focus displacement, thermoelastic distortions on the secondary mirror induce
residual aberrations (spherical aberration, astigmatism and coma) which remains below
$0.2''$ for all power ratios. The effect of thermoelastic distortions is larger on the
secondary mirror than on the primary, but still remains
below the diffraction limit.

The heaters-induced thermoelastic distortions of the primary and secondary mirrors
thus have a minor effect on the optical performances of the telescope. This effect is mostly
limited to an easily compensated focus shift effect. Thus, the turbulence production
is the main cause of the sharpness degradation observed when the defrosting power is
too high.

\subsection{The camera box entrance window}

The largest temperature gradient on the optical path occurs between the M2 and M3 mirrors, precisely at the entrance of the camera box. In order to reduce the temperature gradients at these interfaces, we made the following choices. (i) The camera box was split between an upper part containing the main optical components (such as the M3 mirror), and heated to $-20^\circ$C, and a lower part containing the cameras and electronics, and heated to $0^\circ$C. (ii) The entrance window consisted in a double lens separated by dry air. In order to have a relatively uniform PSF across the focal plane and on the CCD, with at the same time minimizing the number of glass interfaces, we chose to use lenses instead of a planar double glass window \citep{Daban+2010}. 

\begin{figure}
   \includegraphics[width=\hsize]{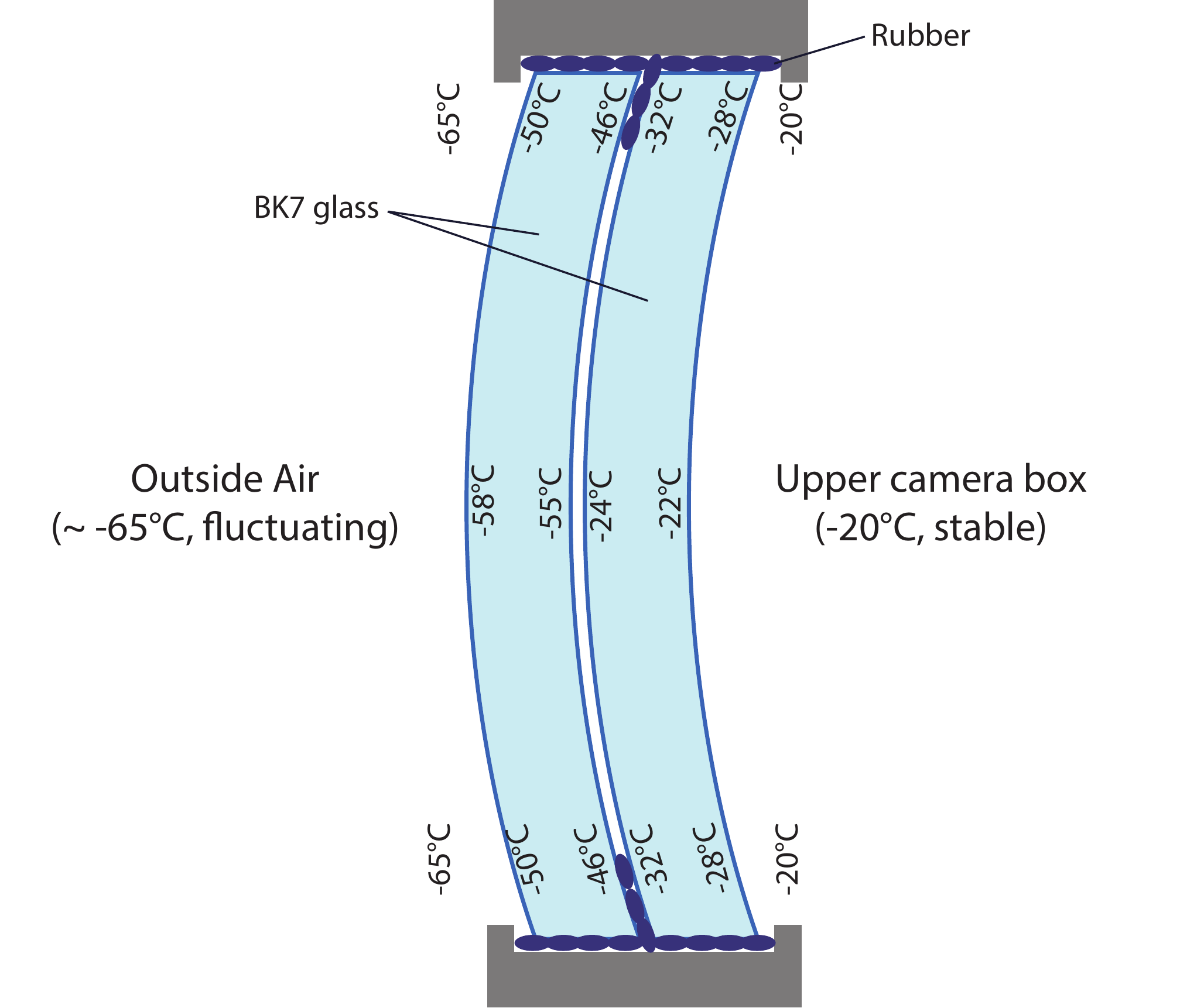}
   \caption{Cross section of the camera box entrance window highlighting some of the material used and expected temperatures for an imposed $-20^\circ$C temperature in the upper camera box and a nominal outside temperature of $-65^\circ$C. }
   \label{fig:lame_astep400}
\end{figure}

Figure~\ref{fig:lame_astep400} shows a cross section of the entrance window which is made of two spherical lenses in crown borosilicate (BK7) glass. Typical temperatures in the air and in the glass are indicated and were calculated using the method described in the Appendix. The advantage of using double glass is that the temperature jump between the outside air and that on the exterior window is only in this case $\sim 7^\circ$C, about half the value that it would have with a single glass window. A large temperature jump of $\sim 30^\circ$C is expected across the layer of air between the glass plates. However, this layer of air is thin, mostly conducting, and is thus not expected to generate a significant amount of turbulence. 

The temperatures obtained in fig.~\ref{fig:lame_astep400} were calculated using the one-dimensional approach described in the Appendix. They account for the fact that rubber is a better heat conductor than air which yields a radial temperature gradient in each lens. A temperature gradient also exists within the lens along the optical path. Given the simplifications, these are only approximative estimates, but they are useful to predict the sign and magnitude of the variations. The main consequences of outside temperature variations are to yield a change in the lens curvature radius of about $17\rm\,\mu m\,K^{-1}$. However, because of the presence of other lenses on the optical path, the variation of the position of the focal plane is expected to be smaller. Furthermore, given our constraint of PSFs spread over at least 2 pixels on the CCD, deformations of the PSF to radial changes of the curvature radius of the lenses may be neglected.

Inside the camera box, small fans homogenize the temperature in the two main areas, the upper and lower boxes in order to minimize temperature fluctuations on the optical path. Turbulence inside the box is therefore expected to have a limited effect on the PSF size. The temperature gradient between the upper and lower parts is stable and hence does not affect dynamically the position of the focal plane.

\subsection{Turbulent seeing estimates}

We now turn to the estimation of turbulent seeing, i.e., the magnitude of the perturbations of the PSF due to turbulence on the optical path. 

Temperature variations on the optical path lead to variations of the index of refraction which affect the wave front. Its variance $\sigma$ can be estimated using \citep{Dalrymple+2004},
\begin{equation}
\sigma^2= 2 K_{\rm GD}^2 \int_0^L \meanrhosq l_z dz, 
\label{eq:sigma}
\end{equation}
where $K_{\rm GD}\approx 0.22\rm\,cm^3/g$ is the Gladstone-Dale parameter which links air refractivity at visible wavelengths to its density, $\rho'$ is the fluctuating density, $l_z$ the correlation length along the optical axis and $L$ is the total path length through the disturbance. \cite{Dalrymple+2004} further discuss that the value of the seeing created by this turbulence depends on whether it is weak (when its variance is larger than the observation wavelength) or strong (otherwise). We focus on the latter, which represents an upper limit. The turbulent seeing $\turbseeing$ is then estimated from the blur angle for $50\%$ of the encircled energy
\begin{equation}
\turbseeing=4 (\ln 2) {\sigma\over l_z}.
\label{eq:theta}
\end{equation}

By using mean values for $\meanrhosq$ and $l_z$ in eq.~\eqref{eq:sigma} and using eq.~\eqref{eq:theta}, we obtain
\begin{equation}
\turbseeing\approx 4\sqrt{2}(\ln 2)K_{\rm GD}\sqrt{{L\over l_z}\meanrhosq}.
\end{equation}
Experiments for mirrors show that the correlation length is generally about $10\%$ of the length of the disturbance and that similarly, the density fluctuations in the air amount roughly to 10\% of the total density variations in the flow. We thus chose to write $L\equiv \xi_L 10 l_z$ and $\meanrhosq^{1/2}\equiv \xi_\rho 0.1 \rho \Delta T/T$, where $\Delta T$ is the temperature difference between the mirror surface and ambient air and $\xi_\rho$ and $\xi_L$ are constants expected to be of order unity. The mirror seeing can therefore be estimated to be
\begin{equation}
\turbseeing\approx 0.28 \xi_\rho \sqrt{\xi_L} (\Delta T/1\,{\rm K})\,{\rm arcsec},
\label{eq:turbseeing_estimate}
\end{equation}
where we have assumed $T=-65^\circ$C and $P=600\,$mbar as appropriate for Concordia, but the same value would be obtained at sea level and $T=20^\circ$C. We note that, with these hypotheses, and since we expect $\xi_\rho\approx 1$ and $\xi_L\approx 1$, the mirror seeing is independent of mirror size and comparable to experimental measurements on large telescopes \citep{Lowne1979, Racine+1991}. 

We expect this estimate with $\xi_\rho\approx 1$ and $\xi_L\approx 1$ to apply for free convection cases, i.e., mirror seeing both due to M1 and M2 and to the entrance window. However, while the size of the disturbance, $L$, is generally similar to the mirror or lens diameter in the case of a horizontal surface, it decreases when this surface is tilted because heat will be transported upward against gravity rather than on the optical path. On the other hand, we expect $l_z$ to be independent of the surface orientation. This implies that we should expect $\xi_L$ to be smaller than unity for the camera entrance window thus reducing perturbations to the PSFs. Similarly, inside the double glass, optical rays cross a distance $L$ equal to the thickness of the layer between the two lenses (5mm), smaller than the expected correlation scale of convective cells $l_z$. We therefore expect a significant reduction of $\turbseeing$ for that case.

\section{The spring-time observations}

We now turn to the analysis of the observation campaign of spring 2013 at Concordia. This campaign was focused on quantifying the sources of PSF broadening in ASTEP~400.

\subsection{Setup}

Observations of Canopus (RA: 06:24:17.5, DEC:-52:43:5.1) were conducted with ASTEP~400 (equipped with an optical density plus an H$\alpha$ filter) between November 17, 2013 and December 10, 2013, i.e. during the Antarctic spring and in broad daylight. The outside temperatures varied from -46$^\circ$C to -25$^\circ$C. Joint observations with a DIMM telescope to measure the atmospheric seeing were performed starting on November 21, 2013. The DIMM was first set up on a platform located at about 6 meters above the ground until on November 25, 2013 it was set up at ground level, in order to obtain turbulence levels comparable to those experienced by ASTEP. 

ASTEP~400 is located in a dome which does not moves azimuthally and with two retractable panels on the North and South, respectively. Given the fact that the Sun never sets at this latitude and in this season, this implied that the telescope was fully in the shade of the dome only twice per day, i.e. from about 8:00 to 12:30 and from 20:30 to 01:00. We however performed nearly continuous observations and also analyze the consequences for the observations of the presence of direct sunlight on the telescope. Although one may think that they bear little evidence for night-time observations, they in fact inform us on the behavior of the telescope in the presence of extreme temperature gradients and extreme turbulence levels. They may also be of interest when related to observations of the Sun with similar instruments. 

In order to analyze the spring campaign observations, we combine the data obtained from ASTEP and in particular the measured size of the PSF of canopus (its Full Width at Half Maximum), the DIMM seeing measurements, the temperature and wind parameters obtained from the meteorological station at the Concordia station, and the temperature measured from our sensors. The individual measurements with ASTEP~400 on Canopus correspond to $2\,$sec exposures. They can be analyzed directly, or combined with DIMM seeing measurements. In that case, we use the median of each measurements over one minute.

\subsection{Atmospheric seeing}

The atmospheric seeing at the Concordia station has been studied thoroughly \citep[e.g.][]{Aristidi+2009,Aristidi+2013}. Atmospheric turbulence there is mostly dominated by a boundary layer whose height varies from 0 to about 50 meters
\citep{Trinquet+2008, Giordano+2012}.
In the antarctic spring and summer the ground is progressively heated to a temperature which is just a few kelvins cooler than the maximum temperature reached by the air during the day. As a result, every day, around 16:00 to 17:00 local time (i.e. about 2 hours after the maximum air temperature) the atmosphere reaches an almost perfectly isothermal state. The resulting fluctuations of the density of the air and hence of its refractive index are thus minimized, so that even in the presence of wind, the turbulence on the optical path remains minimal. 

\begin{figure}
\includegraphics[width=\hsize]{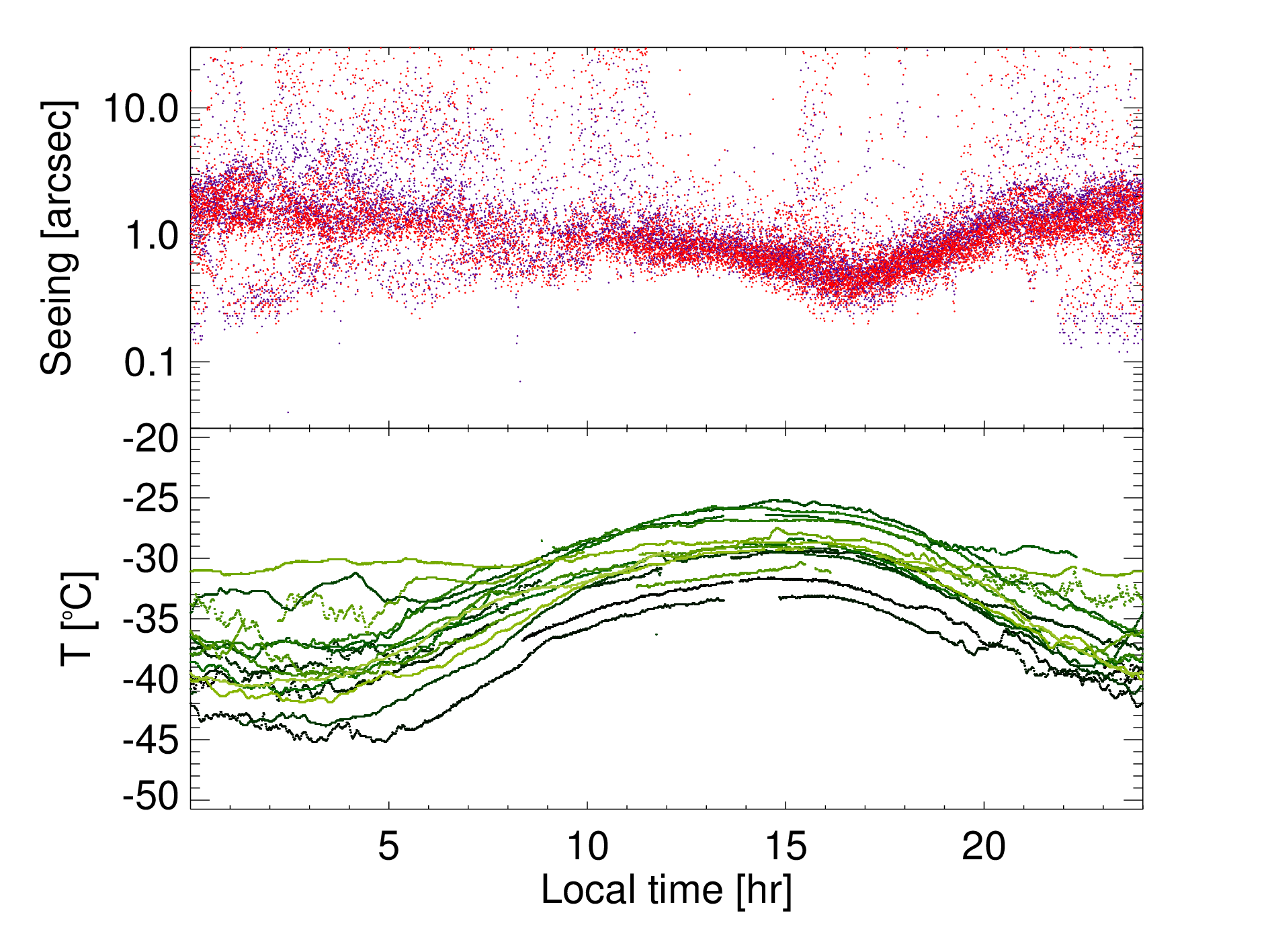}
\caption{Values of atmospheric seeing and atmospheric ground temperature as a function of local time as measured from 21 November 2013 to 9 December 2013. {\em Top panel}: Transversal (red points) and longitudinal (blue) values of the atmospheric seeing measured by the DIMM telescope \citep[see][]{Aristidi+2005}. The low values of the seeing between about 22 and 03 hrs local time correspond to the period when the DIMM was on a 6-meter high platform, i.e. before 25 November 2013. All other measurements were performed from about 1 meter altitude. {\em Bottom panel}: Atmospheric temperature measured at about 1 meter above the ground level. The colors varies from the earliest measurements (black) to the latest ones (pale green). }
\label{fig:hr_seeing}
\end{figure}

Figure~\ref{fig:hr_seeing} shows the ensemble of atmospheric seeing and temperature measurements acquired during the 2013 spring campaign. The slight phase shift between the maximum temperature and minimum seeing is obvious. The seeing measurements are otherwise quite highly variable implying that a precise monitoring is indispensable in order to evaluate the impact of other parameters on the ASTEP measurements. In particular, the fact that the seeing is strongly correlated with the time of the day like other effects such as the dome and baffle seeing (to be discussed in a following section) require simultaneous measurements. 

It is interesting to note that some extremely low values of the seeing (less than $0.2$\,arcsec) correspond to measurements when the DIMM was on the 6-meter high platform, i.e. before its displacement to the ground on 25 November 2013. This highlights the fact that the turbulent boundary layer is often very thin (a few meters high) and is a motivation to seek a higher elevation for the ASTEP telescope.

\subsection{Dome and baffle seeing due to direct sunlight}

\begin{figure}
\includegraphics[width=\hsize]{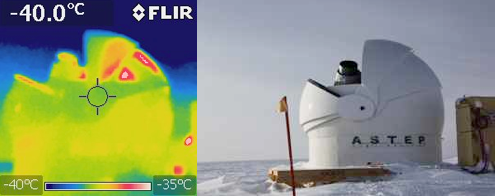}
\caption{Two photographs of the ASTEP~400 dome and upper part of the telescope in the infrared (left) and visible (right), taken on November 26, 2013 at 21:19 while the telescope was in direct sunlight. The temperatures are higher inside the dome (which is black) and also in the upper part of the telescope which receives direct sunlight.}
\label{fig:dome_sunlight_ir_vis}
\end{figure}

The particularity of the spring and summer observations is the presence of the Sun which heats the top of the telescope and the dome, thus bringing a considerable amount of temperature inhomogeneities on the optical path.  As shown in fig.~\ref{fig:dome_sunlight_ir_vis}, the dome illuminated by the sunlight heats up by about 10$^\circ$C even relatively late with a Sun which is only about 25$^\circ$ above the horizon. Being black, the inside of the dome is particularly affected. The telescope baffle is also heated significantly and generates its own turbulence directly above the telescope. 

In order to evaluate the magnitude of the dome seeing, we combine ASTEP and DIMM measurements and write
\begin{equation}
{\cal S}_{\rm Dome}\approx \sqrt{{\cal W}_{\rm ASTEP}^2-{\cal S}_{\rm DIMM}^2-{\cal W}_{\rm intrinsic}^2},
\end{equation}
where ${\cal W}_{\rm ASTEP}^2=({\cal W}_x^2+{\cal W}_y^2)/2$ is the mean FWHM measured in the $x$ and $y$ directions measured by ASTEP, ${\cal S}_{\rm DIMM}^2=({\cal S}_{\rm T}^2+{\cal S}_{\rm L}^2)/2$ is the mean seeing measured in the transverse direction and longitudinally by the DIMM and ${\cal W}_{\rm intrinsic}$ is the intrinsic PSF size of ASTEP below which we cannot go. The latter is estimated from the mimimum FWHM of all measurements at $2.3\,$arcsec. This relatively high value allows spreading the energy over several pixels (the ASTEP pixel size is $0.92$\,arcsec) ensuring a precise photometry \citep{Crouzet+2007}. 

\begin{figure}
\includegraphics[width=\hsize]{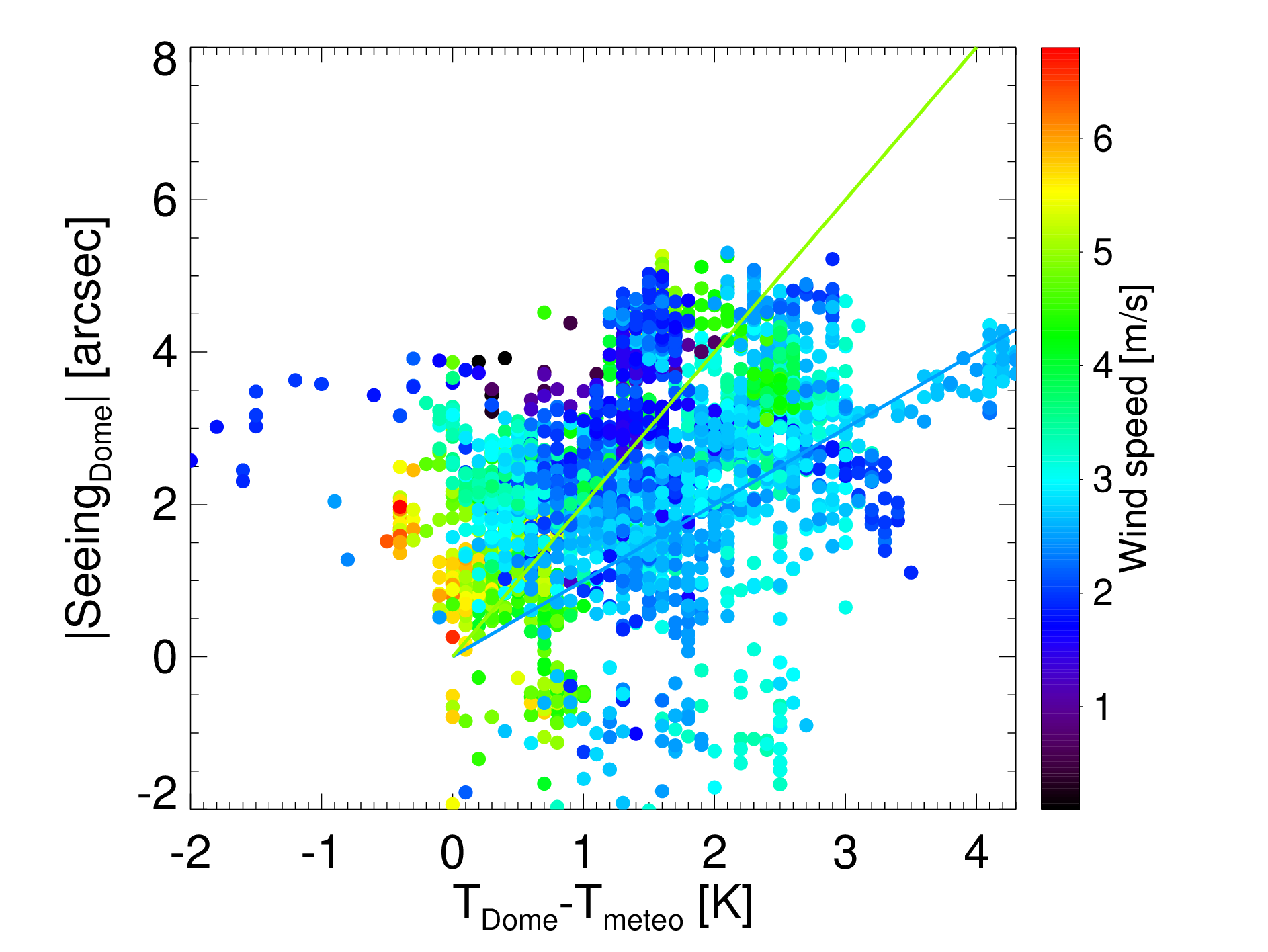}
\caption{Full width at half maximum of the PSF measured by ASTEP~400 as a function of the difference between the measured temperatures in the dome and that given by the meteorological station for a comparable altitude. The points are color-coded as a function of the wind speed, as labelled. The colored curves correspond to approximate fits to the data points for velocities of 2.5 (blue) and 5 m/s (green), respectively.}
\label{fig:tdome_fwhm}
\end{figure}

We show the values of ${\cal S}_{\rm Dome}$ as a function of the difference between the temperature measured from our sensor to the temperature of the meteorological station in fig.~\ref{fig:tdome_fwhm}. In order to obtain this plot, we first verified that when the sun was low or in periods of bad weather, both temperatures were within 1$^\circ$C of each other. We also verified that our guiding was precise to within about half a pixel, so that it would not artificially increase our PSF size in a significant way. Finally, we removed problematic data when the DIMM seeing was above either the ASTEP PSF size or above 3 arcsec, when the weather was too bad (defined as when the median peak flux of Canopus was lower than twice the background flux), and the time periods when we performed other experiments on the instrument (such as heating the mirrors --see hereafter--). Importantly, we noticed some anomalously high values of the temperature of the dome between 15 and 18hrs, local time which corresponded to periods when our sensor was directly illuminated by sunlight. These measurements were also removed from the analysis. 

The resulting fig.~\ref{fig:tdome_fwhm} shows a correlation between the dome to atmospheric temperature difference and the size of the PSFs measured by ASTEP. Because our purpose is not the full characterization of the behavior of our telescope when hit by sunlight, we only crudely analyze this data, but derive an approximate relation between dome seeing, dome to atmospheric temperature difference and wind speed,
\begin{equation}
{\cal S}_{\rm dome}\approx 0.4\,{\rm arcsec} \left(\Delta T_{\rm dome}\over 1\,{\rm K}\right) \left(v_{\rm wind}\over {\rm 1\,m/s}\right),
\end{equation}
where $\Delta T_{\rm dome}\equiv T_{\rm dome}-T_{\rm meteo}$. This expression is approximate to within about a factor of two for a wind speed between 2.5 and 5\,m/s, and it is expected to depend on the particularities of the dome itself. However, it shows that dome seeing is an important factor to consider in the presence of fast temperature fluctuations, especially when the dome's thermal inertia is important. The dependence on wind speed is certainly due to the fact that more wind implies carrying inhomogeneous air on larger distances, thereby increasing the perturbations to the wave front. On the other hand, it is to be noted that wind has another impact, this one positive: it leads to a more efficient cooling of structures and therefore tends to maintain them at temperatures closer to the atmospheric temperature. 
 
\begin{figure}
\includegraphics[width=\hsize]{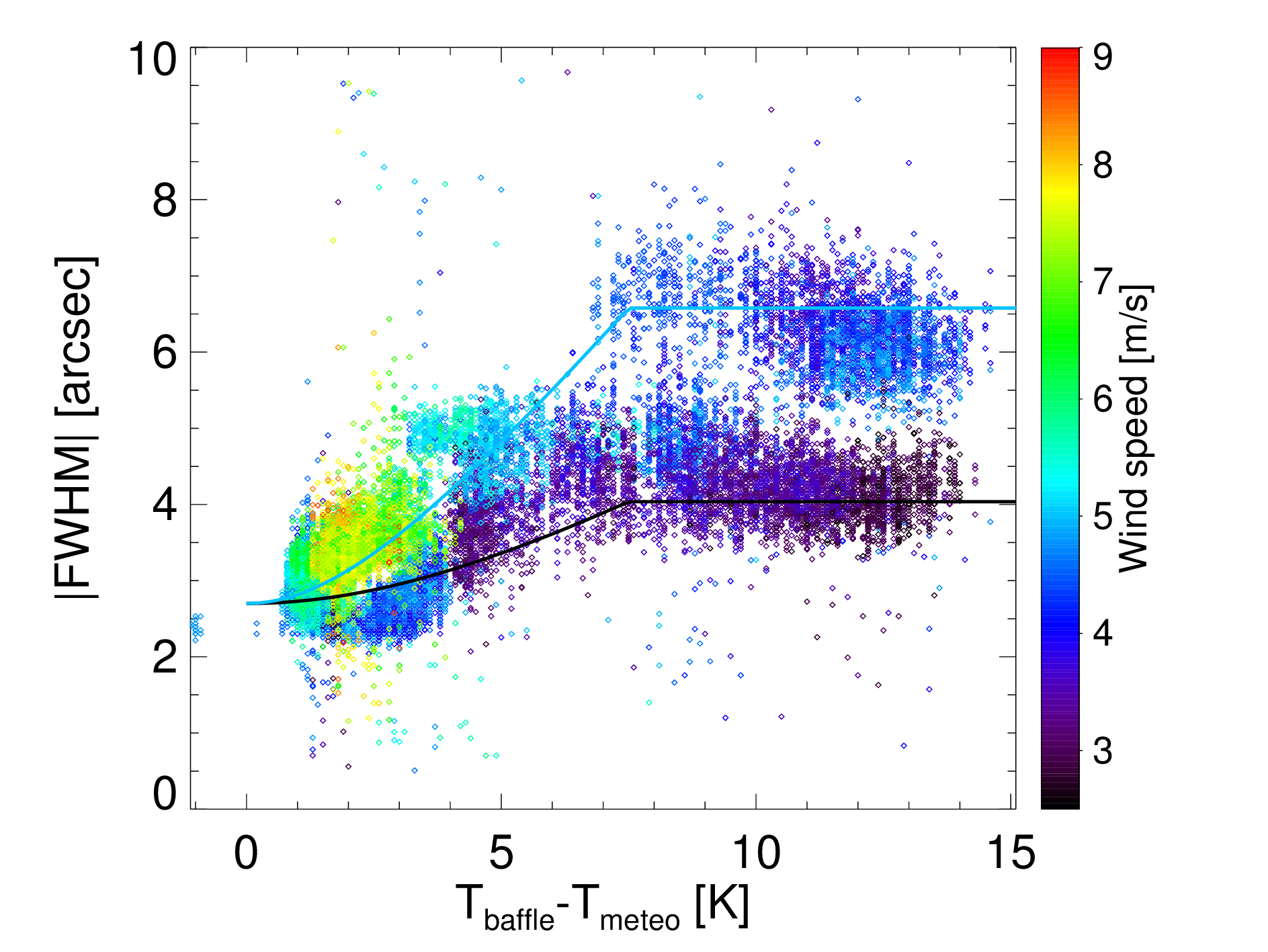}
\caption{Full width at half maximum of the PSF measured by ASTEP~400 as a function of the difference between the measured temperatures in the baffle and that given by the meteorological station for a comparable altitude. The points are color-coded as a function of the wind speed, as labelled. The two black curves correspond to two approximate fits of the data points for low ($\sim 2\rm\,m\,s^{-1}$) and high ($\sim 6\rm\,m\,s^{-1}$) wind speed. }
\label{fig:tbaffle_fwhm}
\end{figure}

Another way to look at the results is through the measurements of the temperature of the baffle of the telescope, which is also directly affected by sunlight and tends to heat up, creating turbulence on the optical part. We chose not to try to separate this effect with that of dome seeing, but present in figure~\ref{fig:tbaffle_fwhm} the values of the FWHM as a function of the difference between the temperature of the baffle and that of the atmosphere. Compared to the previous analysis, we used directly the ASTEP data combined with the temperature measurements, without correcting for variations in the seeing. This is possible in this case because of the larger variations seen on the baffle temperature hit by direct sunlight. Our temperature probe was inside the baffle and protected from direct sunlight so that we did not have to filter for particular moments of the day.  

As in the case of dome turbulence, we see that there are at least two regimes for relatively low wind velocities around 2.5\,m/s and for faster winds around 5 m/s. However, the increase in the perturbation of the wavefront appears to saturate when the baffle becomes warmer than about 7\,K above the ambient temperature. We thus obtain the following simple dependence,
\begin{equation}
{\cal S}_{\rm baffle}\approx {\rm Max}\left[1.2,0.16\left(\Delta T_{\rm baffle}\over 1\,{\rm K}\right)\right]\left(v_{\rm wind}\over 1{\rm \,m/s}\right),
\end{equation}
where $\Delta T_{\rm baffle}\equiv T_{\rm baffle}-T_{\rm meteo}$. Again, this relation is very approximate, but we believe that it is useful as an estimate of the magnitude of these effects. The saturation observed at high $\Delta T_{\rm baffle}$ may be due to the fact that the higher temperature also heats the background so that the temperature fluctuations remain relatively stable. This would take place in a relatively small region of the optical path, given that the telescope was never observing exactly at the zenith. (We could not test this conjecture because no star bright enough to be observed by ASTEP in broad daylight was present at the zenith).

\subsection{Mirror seeing due to M1}

We now turn to experiments directly related to understanding the behavior of the telescope both during the spring and during the cold antarctic winter nights. We first heated the M1 mirror significantly higher than the ambient temperature in order to see the degradation of the PSF due to convection generated inside the tube of the telescope, directly in the optical path. This heating of the mirror mimics the situation that occurs in the winter when cold weather sets in so that the atmospheric temperature drops much more rapidly than the mirror due to its relatively high inertia. Mirror heating is also important to prevent frosting, and it is thus important to estimate the magnitude of this effect. 

\begin{figure}
\includegraphics[width=\hsize]{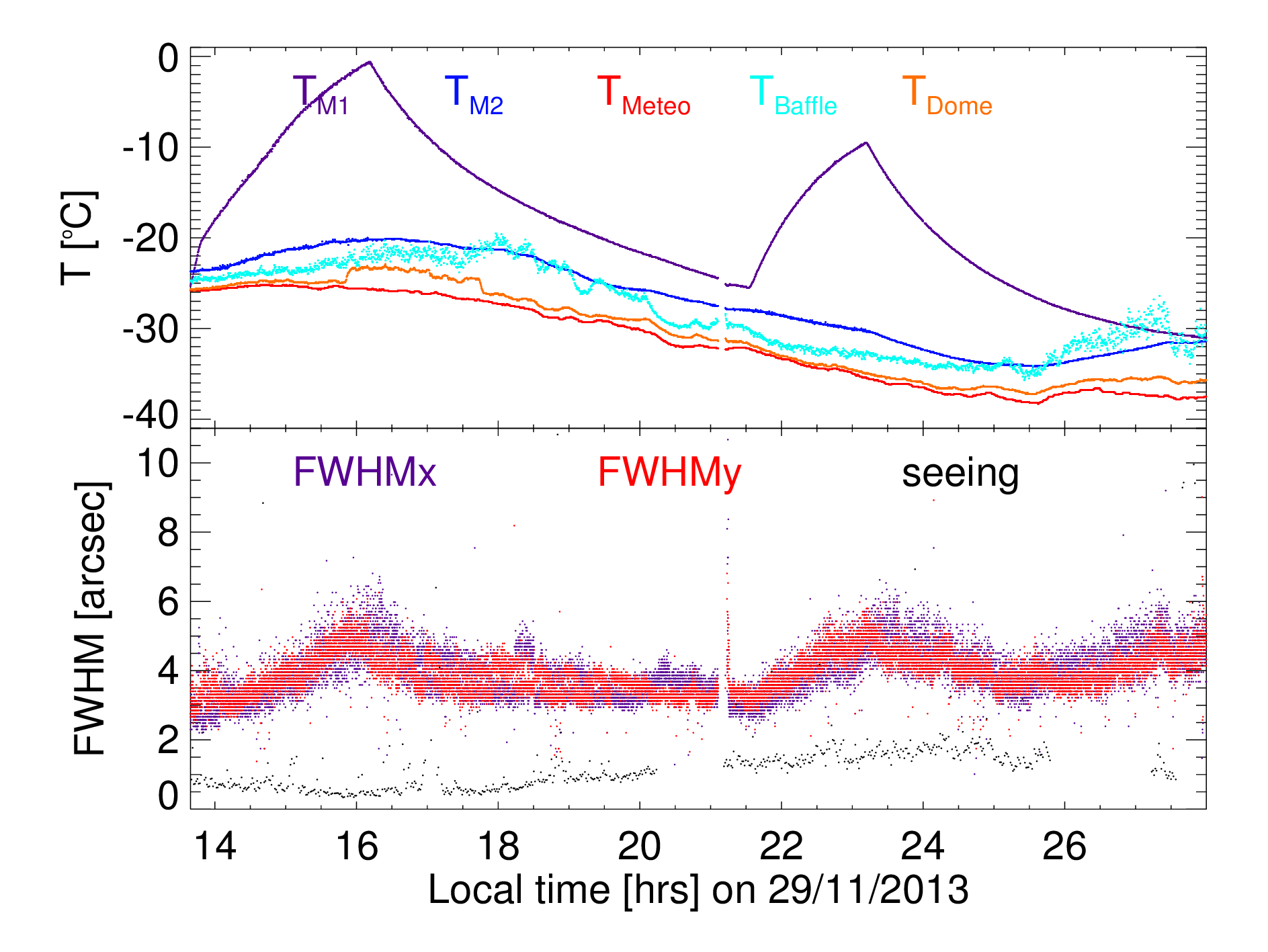}
\caption{Temperatures and FWHM measured during the M1 mirror heating experiments on 29 and 30 Nov. 2013. {\em Top panel}: Temperatures of the M1 mirror (purple), M2 mirror (blue), meteorological station (red), baffle probe (light blue) and dome probe (orange). {\em Bottom panel}: Values of the FWHM in the x (purple) and y (red) directions as measured by ASTEP~400 compared to the atmospheric seeing measured by the DIMM (black).}
\label{fig:tm1_exp1_a}
\end{figure}

We chose a good day characterized by relatively stable temperatures and a good seeing to perform this experiment. As shown in fig.~\ref{fig:tm1_exp1_a}, we increased the mirror heating and let it cool to the ambiant temperature twice. The seeing measured by DIMM was stable and under 2\,arcsec for the entire observation set. The dome and baffle remained within a few kelvins from the ambient temperature and therefore dome and baffle turbulence remained small, except towards the end of the experiment as clearly seen from the increase in measured FWHM. 

\begin{figure}
\includegraphics[width=\hsize]{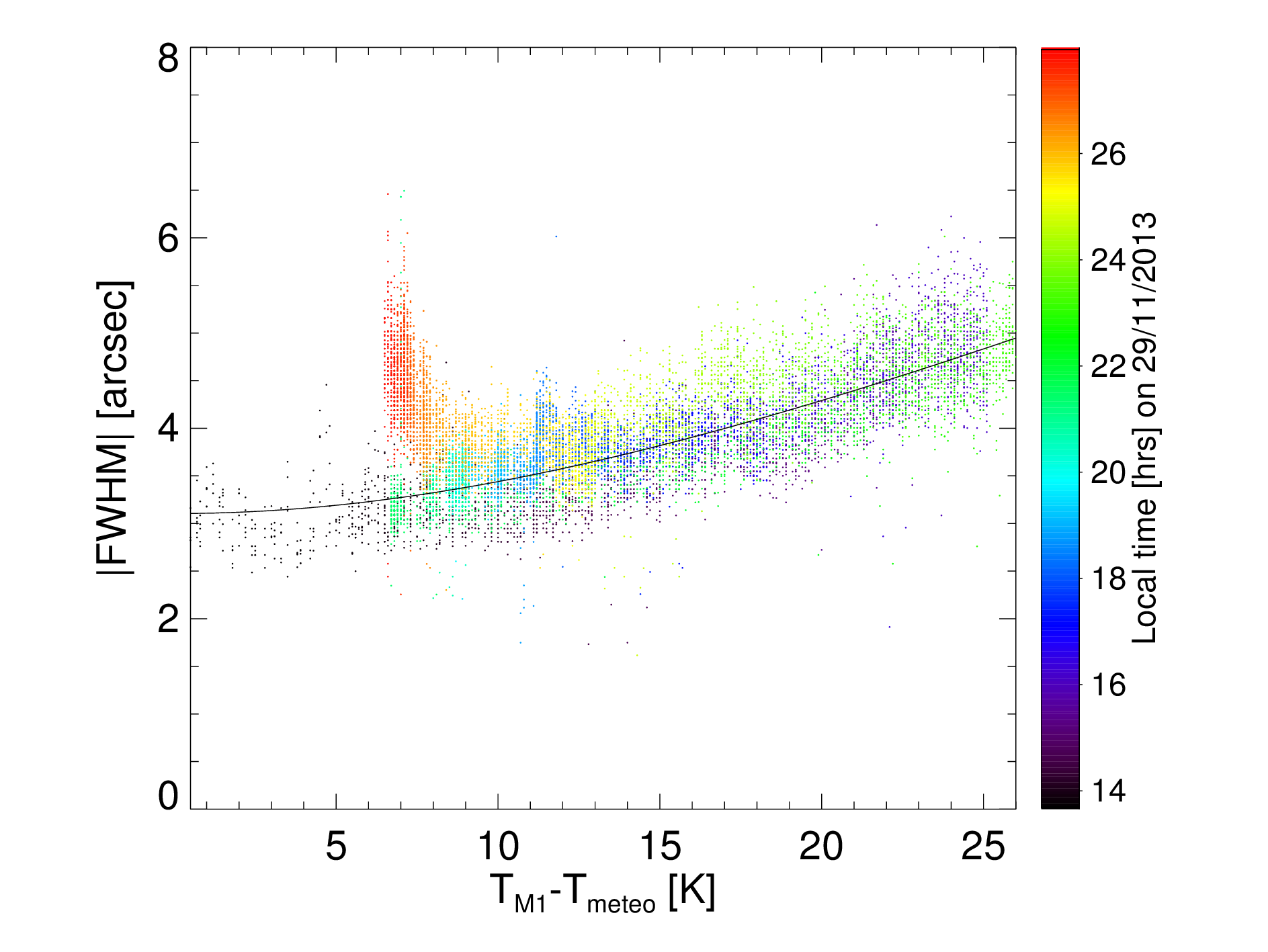}
\caption{FWHM measured by ASTEP~400 as a function of the difference between the temperature of the M1 mirror and that of the atmosphere. The colors indicate the local time at which each measurement was taken. The end of the sequence (in orange and red) is characterized by turbulence due to a high baffle temperature. The black line is a fit to the measurements excluding the ones after 02:00 on the 30/11/2013 [see eq.~\protect\ref{eq:s_m1}].}
\label{fig:tm1_exp1_b}
\end{figure}

Figure~\ref{fig:tm1_exp1_b} shows how the FWHM varies with the difference between M1 mirror and ambiant temperature. We fitted the FWHM data with a function 
\[
{\cal W}=\sqrt{a^2+(b\Delta T_{\rm M1})^2}.
\]
By dropping the points affected by dome seeing at the end of the observation sequence and by weighting as a function of the seeing we obtained $a=3.10$\,arcsec and $b=0.148\rm\,arcsec\,K^{-1}$. Another experiment on 03/12/2013 led to $a=3.23$\,arcsec and $b=0.196\rm\,arcsec\,K^{-1}$. In all these experiments, the telescope angle varied between 38 and 67$^\circ$, but without noticeable effect on the data. We can presume that the convective upwelling plume from the mirror affects the entire telescope tube so that the dependence on telescope angle was weak. 

In summary, we derive an M1 mirror seeing for ASTEP which is
\begin{equation}
{\cal S}_{\rm M1}\approx (0.17\pm 0.03)\,{\rm arcsec}\ \left(\Delta T_{\rm M1}\over 1\,{\rm K}\right),
\label{eq:s_m1}
\end{equation}
where $\Delta T_{\rm M1}\equiv T_{\rm M1}-T_{\rm meteo}$. This is close to the mirror seeing estimated on a theoretical basis in eq.~\eqref{eq:turbseeing_estimate}.

\subsection{Mirror seeing due to M2}

\begin{figure}
\includegraphics[width=\hsize]{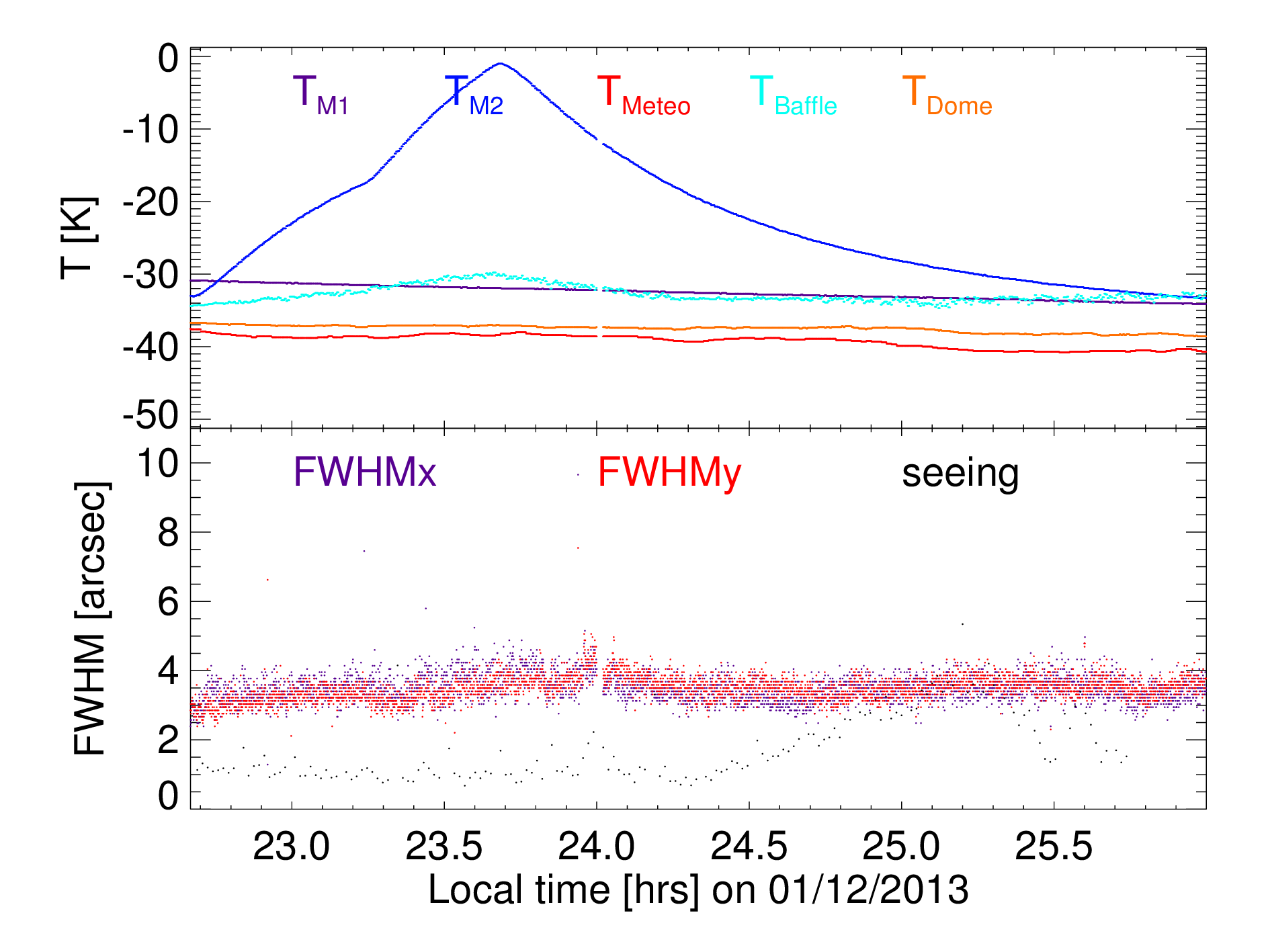}
\caption{Temperatures and FWHM measured during the M2 mirror heating experiment on 01 Dec. 2013. {\em Top panel}: Temperatures of the M1 mirror (purple), M2 mirror (blue), meteorological station (red), baffle probe (light blue) and dome probe (orange). {\em Bottom panel}: Values of the FWHM in the x (purple) and y (red) directions as measured by ASTEP~400 compared to the atmospheric seeing measured by the DIMM (black).}
\label{fig:tm2_exp2_a}
\end{figure}

We performed similar experiments by heating the M2 mirror by up to 37$^\circ$C above the ambient temperature. As shown by fig.~\ref{fig:tm2_exp2_a}, this had a surprisingly small effect on the PSFs measured by ASTEP~400. We interpret this as being due to the fact that the convective plume rising above M2 only intercept a small fraction of the optical path, whatever the direction of the observation.  (M2 is only about 13\% of the surface of M1 when projected in the same plane.) 

As a result, we derived an M2 mirror seeing of
\begin{equation}
{\cal S}_{\rm M2}\approx (0.046\pm 0.03)\,{\rm arcsec}\ \left(\Delta T_{\rm M2}\over 1\,{\rm K}\right),
\label{eq:s_m2}
\end{equation}
where $\Delta T_{\rm M2}\equiv T_{\rm M2}-T_{\rm meteo}$. Note that this M2 mirror seeing is much smaller than the Baffle seeing discussed previously. Indeed, the heating of the baffle leads to a perturbation extending to the entire optical path, hence affecting the PSFs more directly and more severely. 

The change in $T_{\rm M2}$ was also accompanied with a change in focal position of about $-3.5\,\rm \mu m\,K^{-1}$. This corresponds to a dilatation of the structure holding the M2 mirror towards the M1 (hence reducing the M1 to M2 distance) which has to be compensated by a backward motion of the science camera.

\subsection{Additional seeing due to the camera entrance window plume}

\begin{figure*}
\includegraphics[width=\hsize]{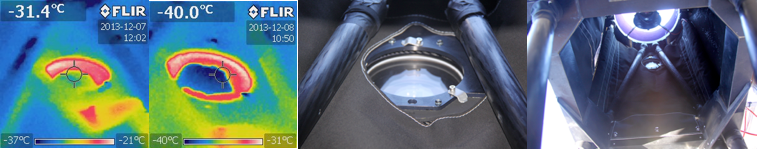}
\caption{Photographs of the camera entrance window in the infrared and visible. {\em First photograph from the left}: Camera box entrance window in the IR when heated to +10$^\circ$C inside the camera box. {\em Second photograph}:  Camera window in the IR when heated -8$^\circ$C (2nd IR image). {\em Third photograph}: Blowup of the camera window in the visible to scale with the IR photographs. {\em Fourth photograph}: Visible photograph is taken from near the M1 mirror, looking up towards the camera window and the M2 mirror.}
\label{fig:camerawindow}
\end{figure*}

Another location prone to added turbulence because of a relatively large temperature gradient across it is the double lens that forms the entrance of the camera box. This window sees the M2 mirror directly. In order to estimate the amount of added seeing due to the presence of this interface, we varied the temperature of the upper part of the camera box between $-8\celsius$ and $+10\celsius$. As shown by fig.~\ref{fig:camerawindow}, the IR photographs indicate that the outside of the window was approximately at a temperature of $-20\celsius$ and $-30\celsius$ for these two situations, respectively.  This corresponds to a temperature gradient between the M2 mirror and the camera window of about $20\celsius$ and $10\celsius$, respectively. 

\begin{figure}
\includegraphics[width=\hsize]{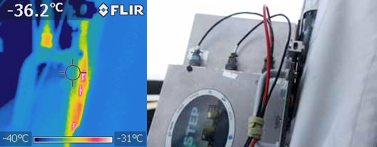}
\caption{Photographs of the ASTEP camera box at the interface with the telescope. {\em Left}: IR photograph showing the warm plume at the interface between the camera box and the telescope tube. {\em Right}: Visible photograph of the camera box and telescope tube.}
\label{fig:camera_and_plume}
\end{figure}

Figure~\ref{fig:camera_and_plume} shows another view of the interface between the camera box and the telescope tube, from which one can clearly see the narrow hotter region, which is bound to generate an upwelling convective plume. 

\begin{figure}
\includegraphics[width=\hsize]{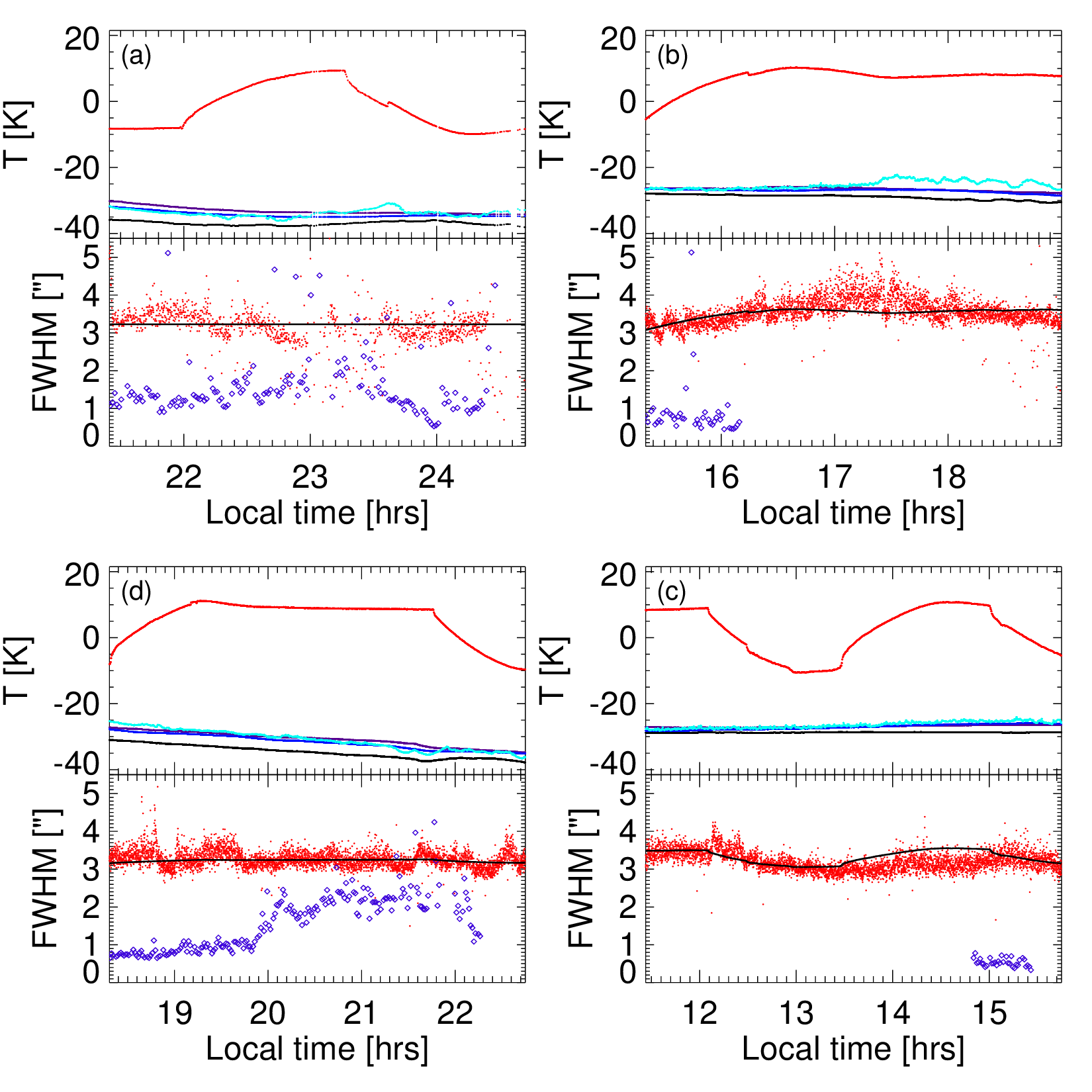}
\caption{Temperatures and FWHM as a function of local time as measured during the camera window heating experiments on (a) 5/12, (b) 6/12, (c) 7/12 and (d) 7/12/2013 (clockwise from the upper left panel). The temperature curves correspond to that of the camera window (red), meteorological station (black), M1 mirror (purple) and baffle (blue). The values of the ASTEP measured FWHM (red) and DIMM seeing (purple) are shown in arcsec. A fit to the FWHM as a function of the difference between the window and meteorological temperatures is shown as a black curve.  The angle of the telescope varied from (a) 60\deg\ to 68\deg, (b) 41\deg\ to 51\deg, (c) 39\deg\ to 42\deg and (d) 49\deg\ to 65\deg.}
\label{fig:twindows_fwhm}
\end{figure}

We present the results of our four camera window heating experiments in fig.~\ref{fig:twindows_fwhm}. These experiments were conducted when the mirror temperatures were stable and close to the ambient temperature given by the meteorological station. The first experiment on 5/12/2003 was affected by clouds between about 23:00 and 23:30. For the other experiments, Canopus was always visible, although some high clouds were present. 

Two experiments (panels (a) and (d)) show no effect of the window heating on the FWHM. Two others ((b) and (c)) show a small but significant increase of the FWHM upon heating the camera window and decrease when cooling it. The experiments showing no noticeable effect correspond to telescope angles above 45\deg, whereas significant effects on the FWHM correlated with the window heating/cooling are only seen when the telescope angle is below 45\deg. This can be interpreted as being due to heat from the window escaping more easily from the telescope tube when it is looking up.  

Quantitatively, we fitted the effect of the heated window from the observed points assuming a constant PSF size for the entire experiment (but allowing it to vary from one experiment to the next) and an additional contribution due to turbulence. As can be seen from fig.~\ref{fig:twindows_fwhm}, the fit only provides a relatively rough estimate of the effect. Evidently, describing the full phenomenon would require a treatment beyond the scope of the present study. Given that caveat, for the cases (b) and (c), our solutions for the turbulent seeing due to window heating are
 \begin{equation}
{\cal S}_{\rm window}\wig{<} (0.11\pm 0.01)\,{\rm arcsec}\ \left(\Delta T_{\rm window}\over 1\,{\rm K}\right),
\label{eq:s_window}
\end{equation}
where we defined $\Delta T_{\rm window}$ as the temperature difference between the outside of the window and the ambient air. Based on our theoretical calculations and direct IR measurements (see fig.~\ref{fig:camerawindow}), we estimated it from the temperature measured on the {\em inside of the window}, $T_{\rm window}$, as $\Delta T_{\rm window}\approx (T_{\rm window}-T_{\rm meteo})/2$. 

\subsection{Forced convection with fans}

The use of fans to limit self-convection is thought to be good to decrease temperature inhomogeneities and hence variations of the air's refractive index. We performed several limited experiments with fans, both on M2 and the camera window and on M1. The first positive effect of fans is to reduce  the temperature difference between the concerned part and the ambient air. The second effect is to prevent convective plumes from rising from hot places into the optical path. 

\begin{figure}
\includegraphics[width=\hsize]{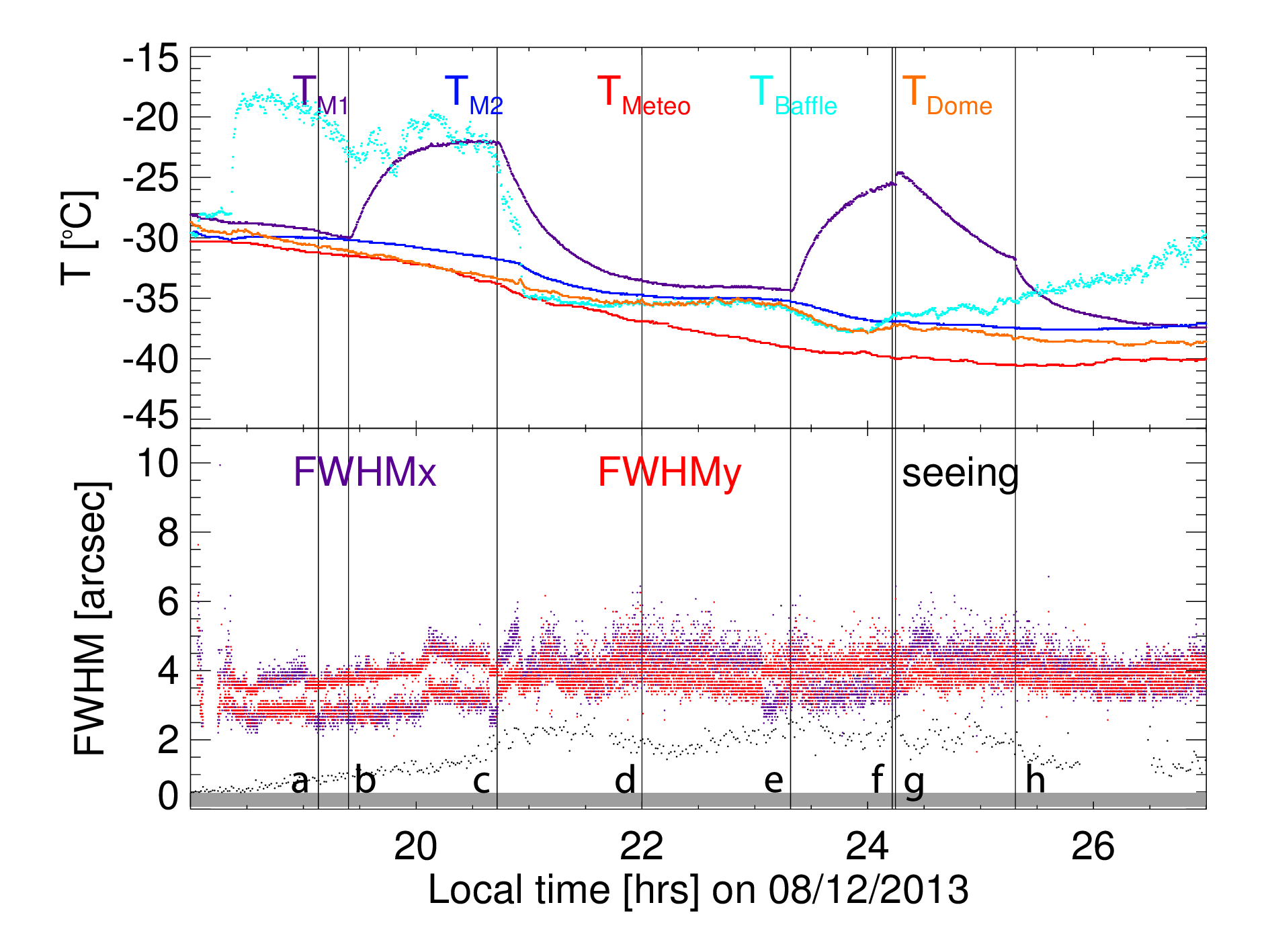}
\caption{Temperatures and FWHM measured during the experiments with a fan over the M1 mirror on Nov. 8, 2013. (see fig.~\protect\ref{fig:tm1_exp1_a} and labels for the meaning of the different colors). The vertical lines indicate the following events (from left to right): (a) Fan on, (b) M1 heating to 100\%, (c) M1 heating to 0\%, (d) moved dome, (e) M1 heating to 100\%, (f) M1 heating to 0\%, (g) fan off, (h) fan on.}
\label{fig:fan_exp}
\end{figure}

Some preliminary experiments with a fan blowing air over the M1 mirror were conducted on the last days of the 2013 spring campaign. Figure~\ref{fig:fan_exp} shows some of the results. Unfortunately, poor weather conditions and some instrumental problems meant that the base PSF was large which prevented quantifying the effect of the fan on the PSF size. Nevertheless, the experiments showed that the use of a fan yields a cooler peak temperature for a given heating power and a much faster cooling of the mirror. The latter can be directly see on figure~\ref{fig:fan_exp} by comparing the cooling without fan from 24:14 to 25:18 (equivalently 00:14 to 01:18 on 09/12/2014) to the one with a fan from 25:18 onward. One can also see that the PSF size increased when the fan was off even though the mirror was cooling, whereas it decreased when the fan was on. This shows that the use of fans should be considered for telescopes in Antarctica. Ideally, air should be taken at the ground level where the temperature (and therefore absolute humidity) is lowest before being blown over the mirror at the ambient temperature there, or a slightly higher temperature.

\subsection{Global dilatations of the telescope}

Whether during day-time or night-time observations, temperature fluctuations yield a global change of the length of the telescope and hence of its focal point. ASTEP~400 is equipped with a piezomotor stage from miCos\texttrademark\ implying that its focal position can be tuned with an accuracy of a few microns. Given that ASTEP~400 is slightly astigmat (which ensures that the PSFs are always spread on at least 2.5 pixels in FWHM), the PSF size and form is relatively constant ($<10\%$ relative change) for a location of the focus of $\pm 20\,\mu$m around the ideal location. However, the large temperature variations implied variations of this focal position of hundreds of microns, leading us to improve on an automatic search for the best position. 

\begin{figure}
\includegraphics[width=\hsize]{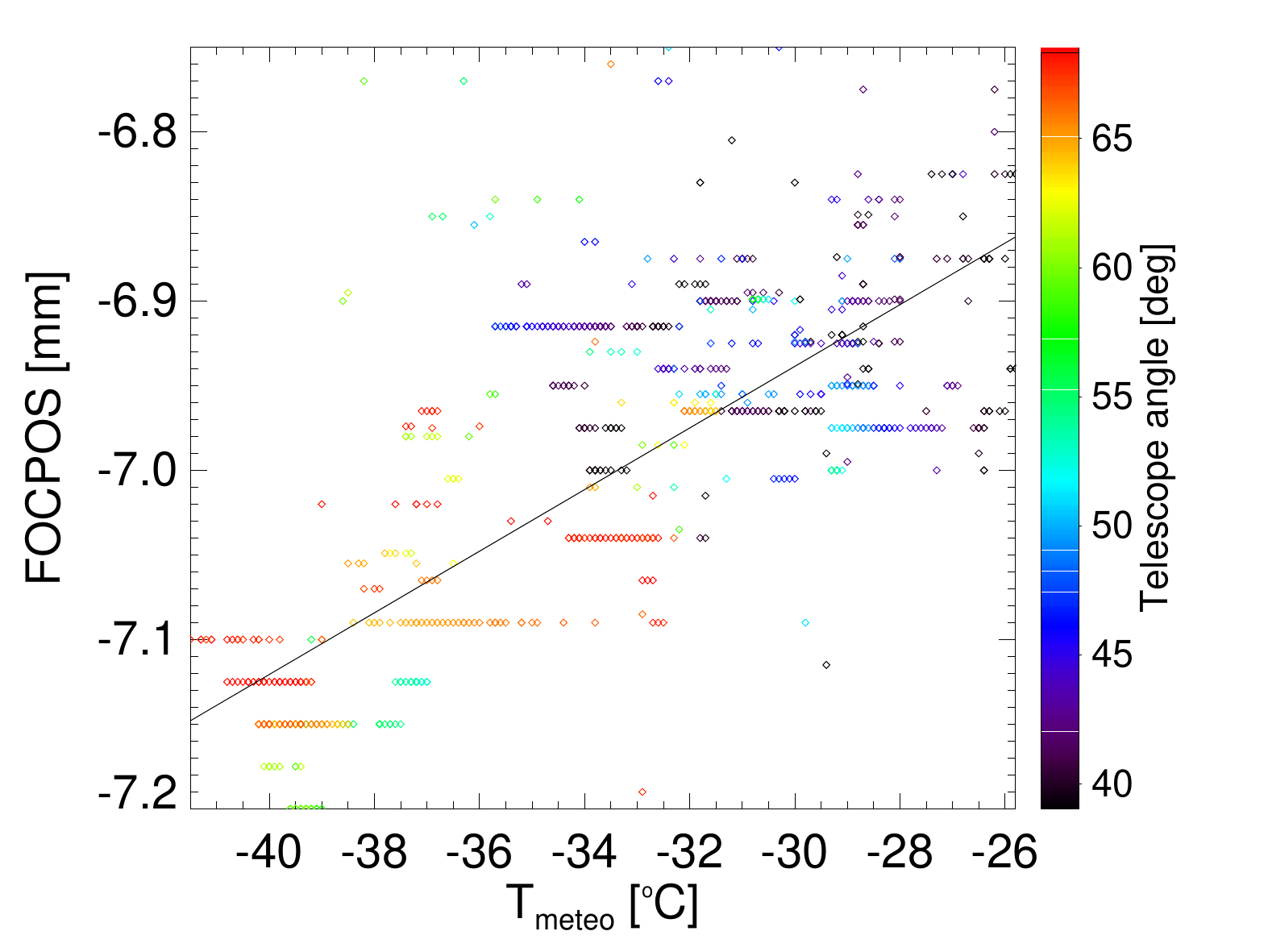}
\caption{Position of the focal point of the ASTEP~400 telescope in millimeters as a function of the outside temperature. The telescope angle is indicated by the color points (no dependence with the focal position is found). The linear regression fit is indicated by a black line. }
\label{fig:tmeteo_focpos}
\end{figure}

The location of the focal plane thus depends on the temperature of the environment and of the telescope's various subsystems. 
Figure~\ref{fig:tmeteo_focpos} shows the linear correlation of $22.1\pm 0.5\,\rm\mu m/K$ between the outside temperature and the position of the piezomotor stage.  There is no dependence between the focal location and the telescope angle showing that flexions of the telescope are not an issue here. (The dependence between the angle and the outside temperature is simply due to the fact that the angle is directly related to the local time, which is directly correlated to the outside temperature.)

This dependence of the focal point has multiple origins: The first, most direct one is the dilatation of the telescope itself. Given the focal length of $2\,$meters, this implies a dilatation coefficient of $11\times 10^{-6}\,\rm K^{-1}$. For comparison, the expansion coefficient of aluminum is $23\times 10^{-6}\,\rm K^{-1}$ and that of the carbon fibers which form the structure of the tubes is about $2\times 10^{-6}\,\rm K^{-1}$.  Another source of variation of the focal position is a change of the curvature radius of the optical systems. A global change of temperature in the M1 mirror will change its curvature radius and hence the locus of its focal point by a mere $-0.3\,\rm \mu m\, K^{-1}$. As discussed in Sect.~\ref{sec:M1}, a vertical temperature gradient (either because of heating or an outside temperature change and the mirror's thermal inertia) can change the focal plane by a greater extent of order $-3\,\rm\mu m/K$. But a more important change is expected to be due to the change of the curvature radius of the camera box entrance window, which could yield a change of up to $17 \,\rm \mu m\, K^{-1}$ of the location of the focal plane. 

In order to estimate the various causes of the variations of the focal point, we took our entire set of data, filtered out from the periods when the peak flux was not higher than at least twice the background, and fitted a multiple variable linear function,
\begin{eqnarray}
\focpos &\!=\!&a_0 + a_1 T_{\rm meteo} + a_2 \Delta T_{\rm M1} + a_3 \Delta T_{\rm M2} \nonumber \\
&&+a_4 \Delta T_{\rm window}+ a_5 \delta T_{\rm box} +a_6 \alpha_{\rm tel},
\label{eq:focpos_T}
\end{eqnarray}
where $\Delta T_{\rm M1}\equiv T_{\rm M1}-T_{\rm meteo}$, $\Delta T_{\rm M2}\equiv T_{\rm M2}-T_{\rm meteo}$, $\Delta T_{\rm window}\equiv T_{\rm window}-T_{\rm meteo}$, $\delta T_{\rm box}\equiv T_{\rm box}-T_{\rm window}$ and $\alpha_{\rm tel}$ is the pointing angle. 

The results, obtained when adding one variable at a time are presented in Table~\ref{tab:instrmodel}. The reduced $\chi^2$ values were calculated from the predicted and the measured focal positions and an estimated uncertainty of $\pm 30\rm\,\mu m$ on the latter. The standard deviation of the focal position was obtained for each parameter considered by multiplying the standard deviation of the parameter considered by the amplitude of the variation (e.g., in the case of $T_{\rm meteo}$,  $4.5\,\rm K \times 8.58\,\rm\mu m/K=38.9\,\mu m$).

From the full model in Table~\ref{tab:instrmodel}, and by order for the largest standard deviation, we obtain that position of the focal point is most affected by: (1) the temperature gradient inside the camera box, with a rate of $5.7\rm\,\mu m/K$; (2) the outside temperature which yields a global dilatation of the telescope at a rate of  $8.6\rm\,\mu m/K$; (3) the temperature gradient at the camera box entrance window with a rate of $5.6\rm\,\mu m/K$; (4) flexions of the telescope which seem to also affect the locus of the focal point at a rate of $-2.4\rm\,\mu m/deg$ (where the angle is that of the telescope tube measured from the horizontal axis); and (5) the temperature gradient inside the M1 mirror at a rate of $6.5\rm\,\mu m/K$ and. The temperature gradient inside the M2 mirror appears to have a comparatively smaller effect, as expected.

\begin{table*}
\caption{Instrument model for the focal point of the telescope as a function of various parameters}\label{tab:instrmodel}
\begin{tabular}{llllllll}
\hline\hline
\multicolumn{7}{l}{Variables} & $\chi^2$ \\
Cte & $T_{\rm meteo}$ & $\Delta T_{\rm M1}$ & $\Delta T_{\rm M2}$ & $\Delta T_{\rm window}$ & $\delta T_{\rm box}$ & $\alpha_{\rm tel}$ & \\
$[\rm\mu m]$ & $[\rm\mu m/K]$ & $[\rm\mu m/K]$ & $[\rm\mu m/K]$ & $[\rm\mu m/K]$ & $[\rm\mu m/K]$ &$[\rm\mu m/deg]$ &  \\ \hline
-6934.9 &             &                   &                         &                         &                     &                                   & 8.19 \\
-6727.1 & 6.01     &                   &                         &                         &                     &                                   & 7.36 \\
-6656.1 & 8.63     &   4.98         &                         &                         &                     &                                   & 7.22 \\
-6654.3 & 8.71     &   4.92         &   0.38               &                         &                     &                                   & 7.22 \\
-6654.6 & 9.14     &   4.89         &   0.42               &    0.53              &                     &                                   & 7.21 \\
-6668.4 & 12.80   &   6.29         &  -2.17               &   5.65              &   6.22           &                                   & 5.80 \\
-6677.9 & 8.58     &   6.53        &   -2.58             &    5.59             &   5.71           &  -2.41                        & 5.60 \\ \hline
\multicolumn{7}{l}{Standard deviation [$\mu$m]} &  \\ \hline
 & 38.9 & 21.6 & -8.7 & 31.2 & 39.1 & -23.8 & \\
\hline\hline
\end{tabular}
\end{table*}

\begin{figure}
\includegraphics[width=\hsize]{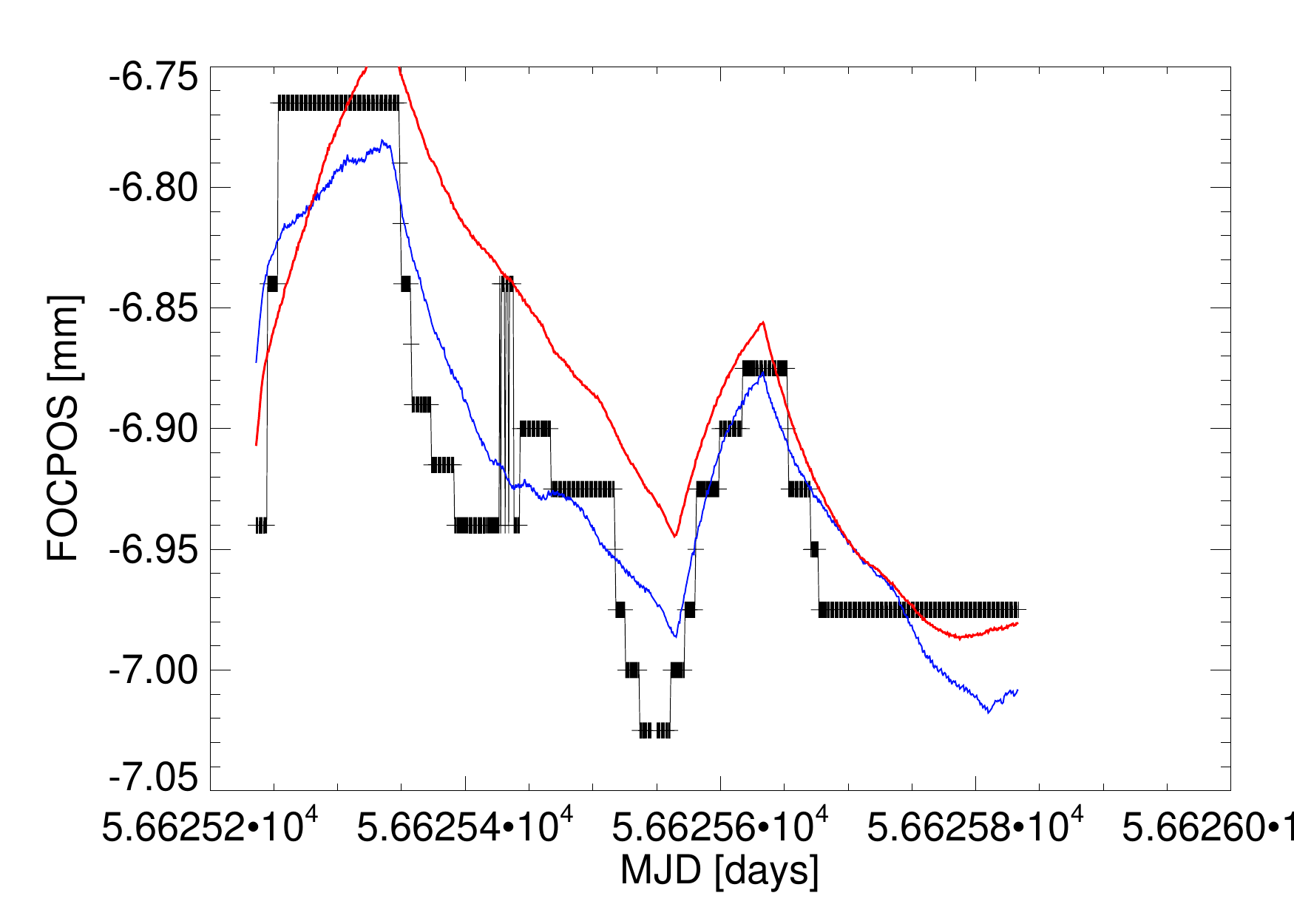}
\caption{Position of the focal point of the ASTEP~400 telescope in millimeters as a function of the MJD date during the first M1 mirror heating experiment. The two episodes of M1 heating are characterized by an increase of the focal point. The black line indicates the measured position of the focal plane. The red curve is the result of the global fit from table~\protect\ref{tab:instrmodel} temperature of the M2 mirror in Kelvins. The blue curve is the result of a fit using only this limited set of data.}
\label{fig:instrmodel_tm1_exp}
\end{figure}

Figure~\ref{fig:instrmodel_tm1_exp} shows the result of the global fit (with the full dataset) applied to one of our M1 mirror heating experiments. Although our global fit (red curve) is obviously not the ideal representation of the behavior of the focus, it reproduces it correctly, with a reduced $\chi^2=2.4$. For comparison a new fit with the same variables for only this heating experiment dataset is indicated with a blue curve. It has a $\chi^2=1.2$ but also an unphysically high value of the effect of $T_{\rm window}$ of $68\,\rm \mu m/K$.  

\subsection{Out-of-focus observations}

The rapidly varying seeing on the ground implies that keeping the instrument in focus is difficult and may be detrimental to the observations. During the summer 2013 campaign, we could use a self-made automatic focusing software. Its principle was based on the slight astigmatism of the telescope that allowed a very direct estimate of the location of the ideal focusing position. However, this software becomes less reliable in bad-seeing conditions and was not used during the winter. An estimate of the consequence of out-of-focus observations is therefore required. 

\begin{figure}
\includegraphics[width=\hsize]{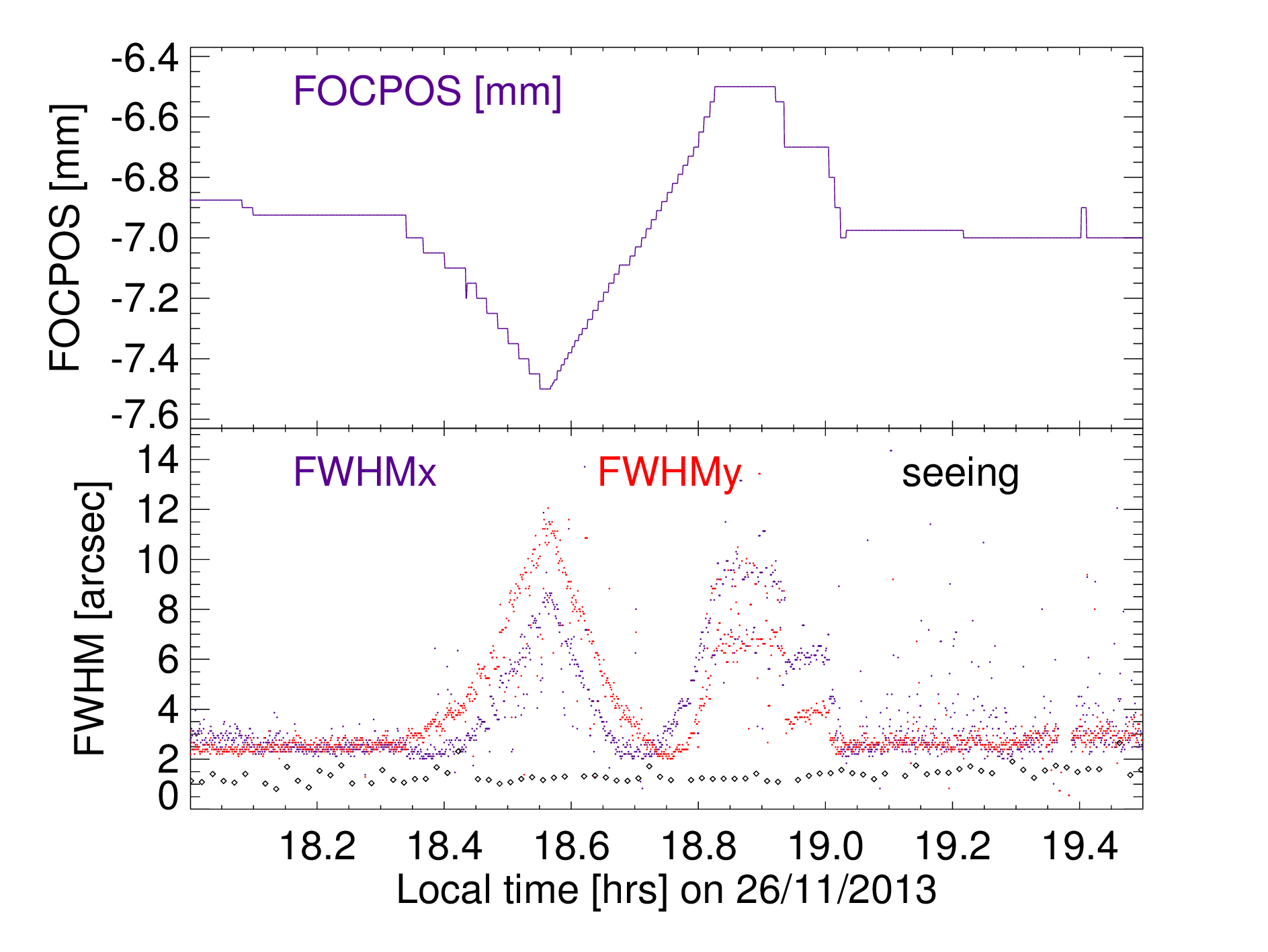}
\caption{{\em Top panel:\/} Imposed position of the focal plane as a function of time. {\em Bottom panel:\/} Resulting full width at half-maximum of the PSF as measured along the x (purple) and y (red) directions, respectively. The seeing measured by the DIMM is indicated by black diamonds. {\em Bottom panel:\/} Values of the FWHM in the x and y directions as a function FOCPOS, the position of the focal plane.}
\label{fig:focpos_exp1}
\end{figure}

In fig.~\ref{fig:focpos_exp1}, we report the result of one experiment done during the 2013 summer campaign, in which we turned the automatic focusing off and forced a variation of the position of the focus while measuring the size of the PSF in the $x$ and $y$ directions. The fact that the FWHM is different in the two directions is a consequence of the slight astigmatism of the telescope. 

\begin{figure}
\includegraphics[width=\hsize]{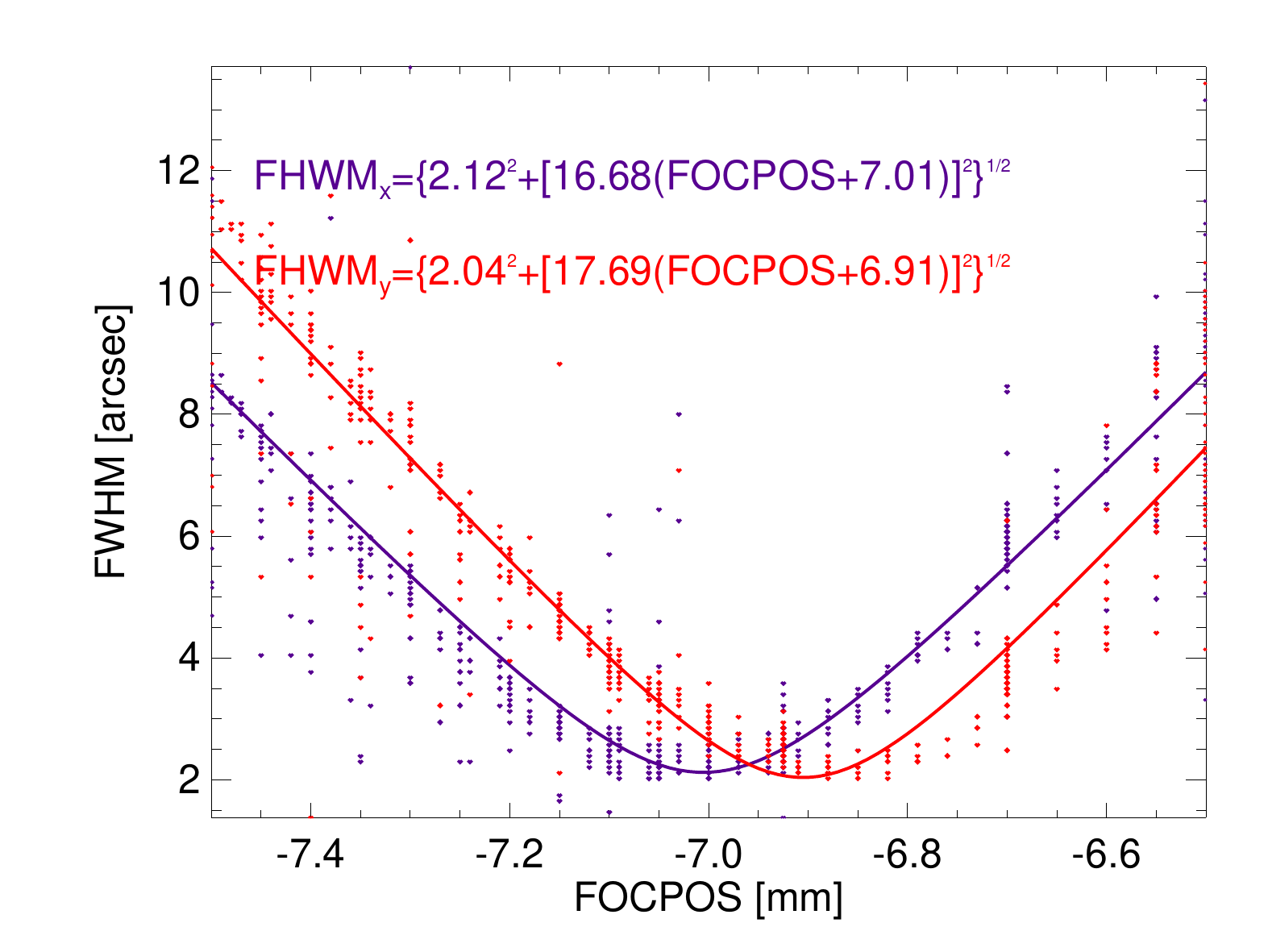}
\caption{Values of the FWHM in the x and y directions as a function FOCPOS, the position of the focal plane obtained in the out-of-focus experiment on 26/11/2013 (see fig.~\protect\ref{fig:focpos_exp1}). The curves shows the fits to the data (as labelled).}
\label{fig:focpos_fwhm1}
\end{figure}

Figure~\ref{fig:focpos_fwhm1} shows the same data but with the FWHM $\fwhm$ as a function of $\focpos$, the position of the focal plane. The ideal focus position is different in $x$ and $y$, again as a consequence of astigmatism. The measurement show a very clear linear variation between $\fwhm^2$ and $\Delta\focpos^2$ (the displacement from the optimal focus) that can be fitted with the relation
\begin{equation}
{\fwhm}\approx \left\{{\fwhm}_0^2 +\left[a \Delta\focpos\right]^2\right\}^{1/2}.
\end{equation}
As fig.~\ref{fig:focpos_fwhm1}, for both the $x$ and $y$ directions, we estimate $a\approx 0.017\rm \,arcsec\,\mu m^{-1}$ and ${\fwhm}_0\approx 2\,\rm arcsec$. Other experiments carried out during the summer season agree with these estimates.

Eq.~\eqref{eq:focpos_T} implies that the focal plane moves with a variation of the global temperature of the telescope by $\sim 8\,\rm \mu m\,K^{-1}$. This implies that after focusing the telescope a temperature variation will tend to increase the FWHM by about $0.14\, \rm arcsec\,K^{-1}$. In the absence of an autofocusing method during the winter, we expect temperature changes of $20\,$K to yield a quadratic increase of the FWHM by up to $3\,$arcsec. 

\section{The winter observations}

\subsection{Temperature measurements}

\begin{figure}
\includegraphics[width=\hsize]{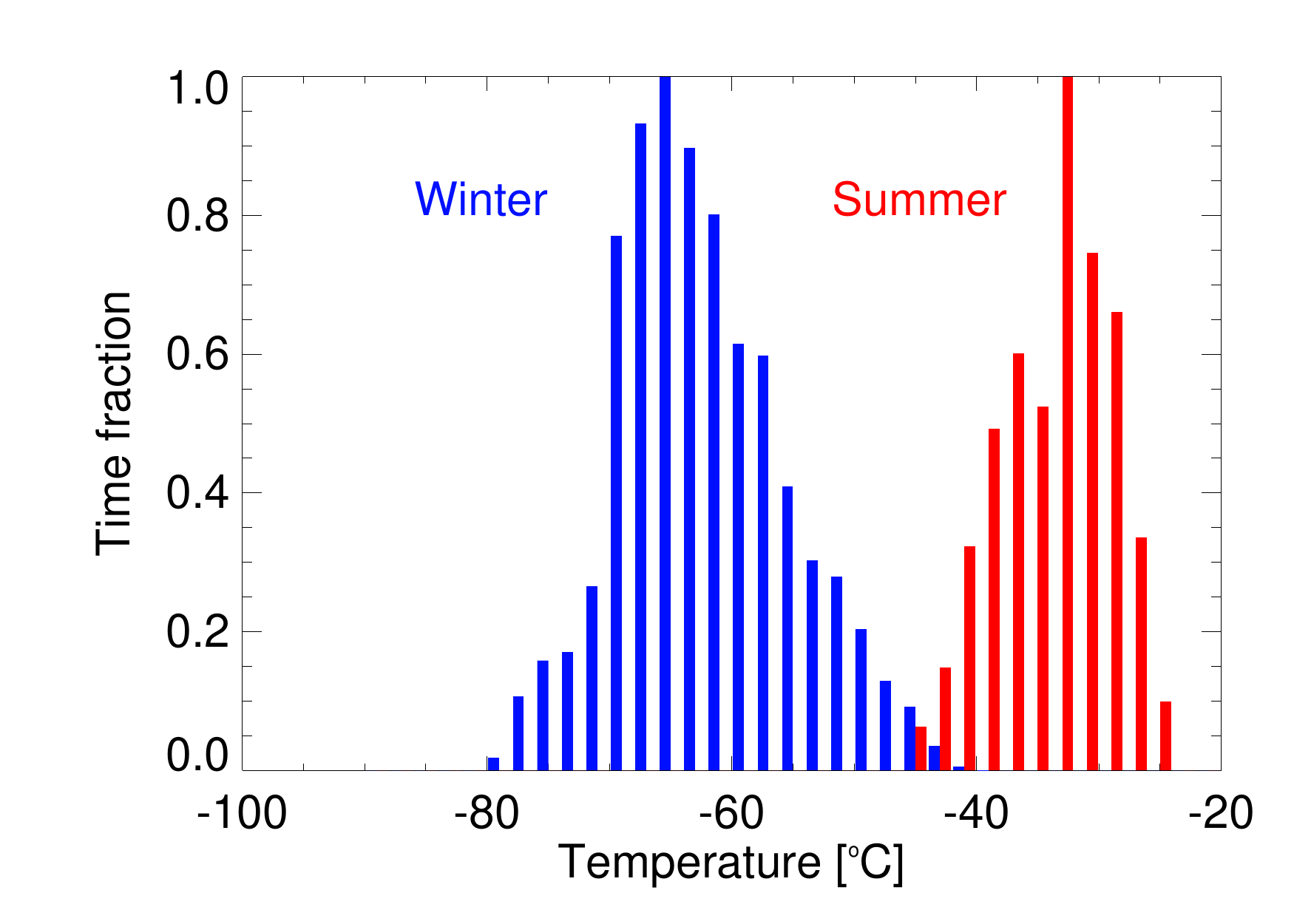}
\caption{Distribution of outside temperatures measured during the two periods during which ASTEP~400 was operating in 2013 labelled ``winter'' (March to September) and ``summer'' (November and December).}
\label{fig:temperatures}
\end{figure}

Operating a telescope at Concordia implies coping both with extremely low and highly variable temperatures. Figure~\ref{fig:temperatures} shows the distribution of outside temperatures measured at about 2 meters above the ground, at about the same height as ASTEP~400, during the telescope operations in 2013, both during the winter and summer campaigns. During the ``summer'' campaign (actually taking place during spring, i.e. from mid-November to mid-December), the temperatures ranged from $-45^\circ$C to $-25^\circ$C. During ``winter'' (March to September), the temperatures ranged from $-80^\circ$C to $-40^\circ$C. The median temperature was $-65^\circ$C. Most of the observations took place in a range between $-70^\circ$C and $-55^\circ$C. 

\begin{figure}
\includegraphics[width=\hsize]{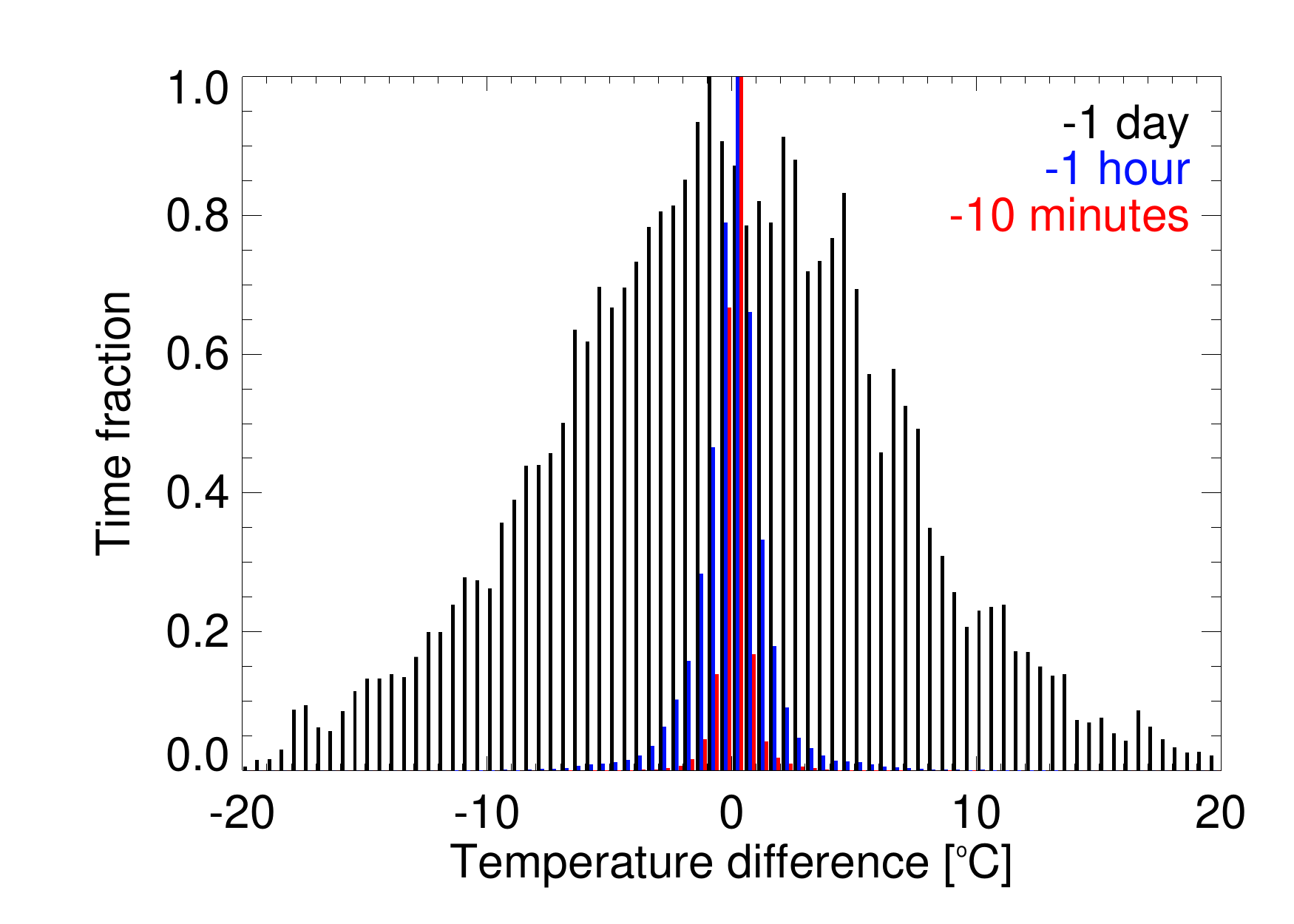}
\caption{Histograms showing the difference in temperature between that measured at time $t$ and 1 day (black), 1 hour (blue), and 10 minutes (red) before, respectively.}
\label{fig:tgradients}
\end{figure}

The temperatures where also rapidly variable. As illustrated in fig.~\ref{fig:tgradients} for the ``winter'' 2013 season only, the day to day temperature could vary by up to $\pm 20^\circ$C, with a standard deviation of $6.9^\circ$C. Over 1 hour, the variations could amount to $\pm 6^\circ$C with a standard deviation of $1.5^\circ$C. Over ten minutes, the temperature fluctuations were still significant, having a standard deviation of $0.61^\circ$C.

These rapid temperature variations have two consequences: (i) Given the thermal inertia of the telescope (in particular of the M1 mirror) which yields relaxation half-times of the order of $\sim 1$ to $2$ hours (see fig.~\ref{fig:tm1_exp1_a}), one can expect the telescope to have a temperature that differs from the outside one by typically one to a few $^\circ$C; (ii) When the telescope is cooler than the outside and the relative humidity high enough, condensation may take place. This required heating the telescope and in particular the M1 and M2 mirrors. 

\subsection{Tracking quality}


\begin{figure}
\includegraphics[width=\hsize]{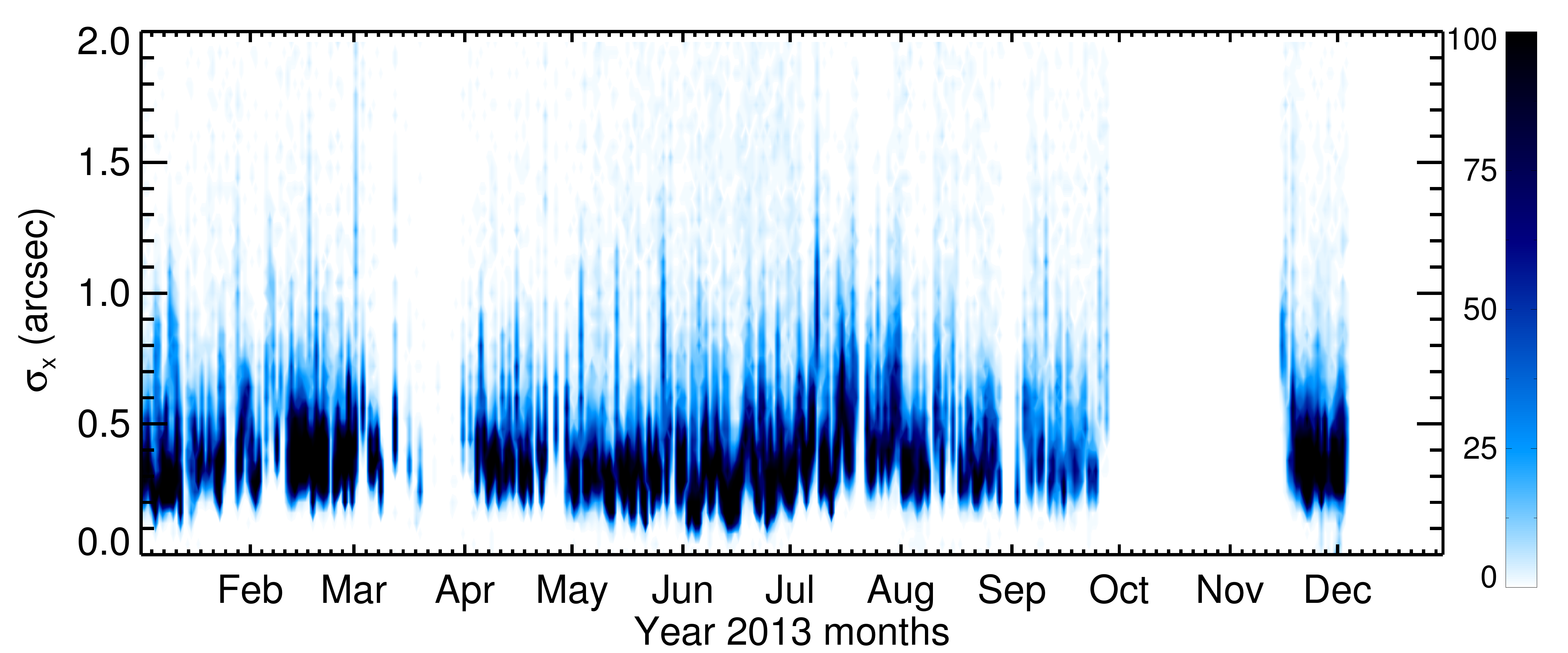}
\includegraphics[width=\hsize]{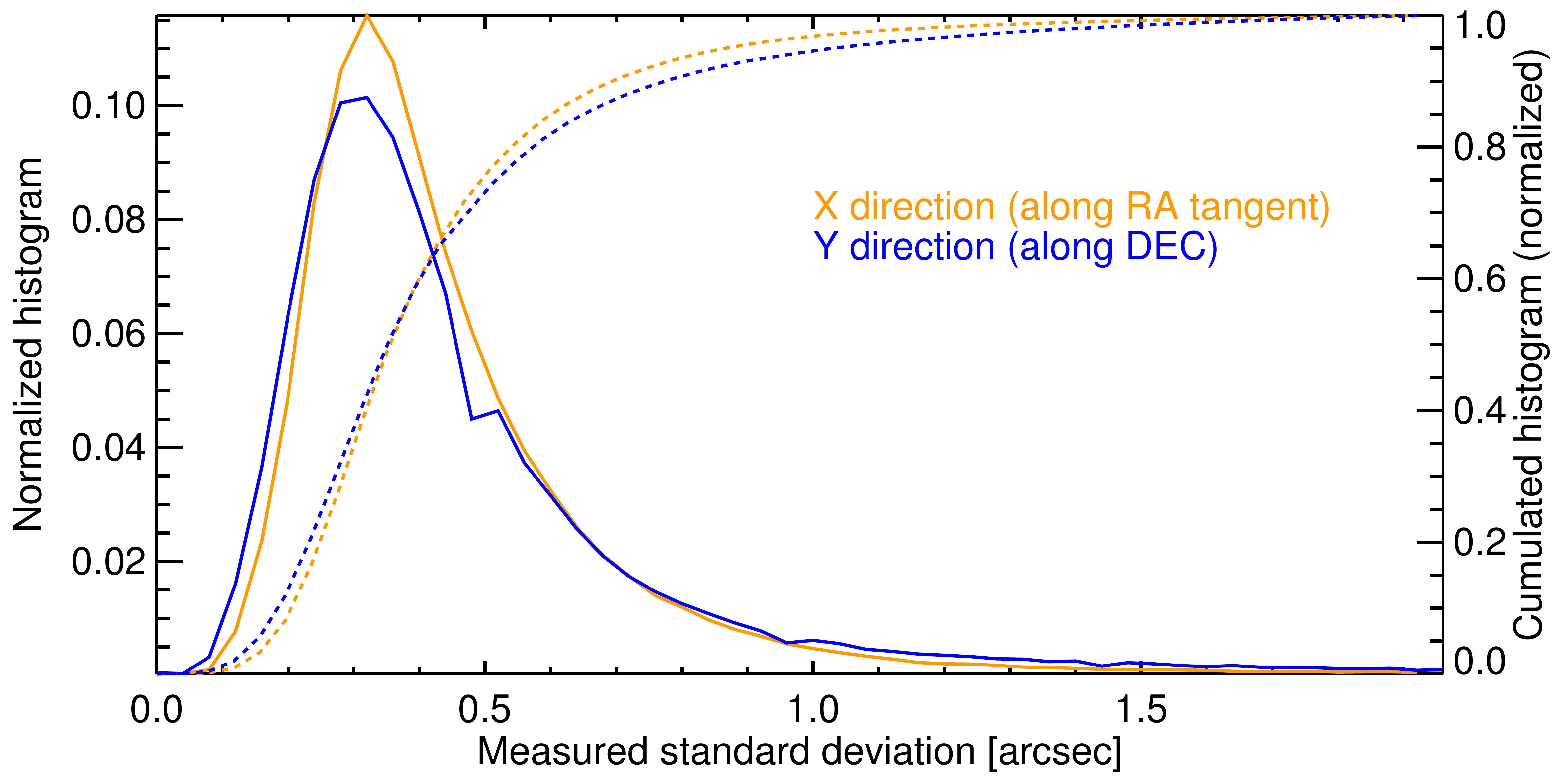}
\caption{\label{fig:guidinghistograms} Telescope guiding statistics for the whole year 2013. (top) The guiding standard deviation was computed over 1 minute bins and then accumulated to obtain one histogram for each day of the year. (bottom) Normalized Average histograms (plain lines) of the guiding standard deviation, and normalized cumulated histograms (dashed lines) for the year 2013.}
\end{figure}

The telescope is guided thanks to a camera using the blue part of the spectrum \citep[see][]{Daban+2010} at a typical frequency and integration time of about 0.3\,Hz and $\sim\!2$\,s respectively. At such low frequency, the guiding provides tracking-drift compensation rather than an adaptive tip-tilt system. The top plot of Fig.\,\ref{fig:guidinghistograms} represents daily histograms of the guiding standard deviation (here in along the tangent to right ascension only), evaluated over 1 minute bins. The bottom plot shows the yearly normalized average histogram (plain lines), and the normalized cumulated histogram (dashed lines) for both guiding directions (tangent to right ascension and approximately along declination). The data related to the testings described in this paper correspond the mid-November to early December period (top plot). The guiding error is found to be below 0.5'' about 80\% of the time with a peak value at $\sim\! 0.3$''. This shows that the guiding poorly account for PSF broadening and mainly compensates low-frequency mount tracking drifts. The results are show here for the year 2013, but similar values were obtained for the previous years (except for the 2010 campaign where we did not have guiding logs).

More significant consequences on the PSF broadening can occur on long-exposures (e.g. over 1 minute) when the mechanical parts (gears) present so-called ``backlash'' that are not correctly compensated for by the driving system (that integrates a backlash compensation option). Variable mechanical backlash occur because of the imperfection of the gear system (that was not optimized for 24 hours tracking) and its evolutions due to temperature changes. The ASTEP\,400 control software includes an automatic backlash estimation and compensation that tries to optimize the backlash parameters: too high values result in saw-tooth guiding curves, while too low values result in rectangular shaped curves with typical amplitudes of $\pm 1$''. But these effects are rare and are rather efficiently compensated for by software when they occur.

\subsection{Seeing and PSF measurements}

We analysed ASTEP~400 images from the two first winter seasons in 2010 (May 26th--Sep. 24th) and 2011 (March 29th--August 12th). For each image a mean FWHM of the PSF was estimated from all detected stars, leading to a total of more than 100000 FWHM values spanning these two winters. These data were compared with seeing values obtained at the same time (within an interval of 2 minutes) by a DIMM located on a 6m high platform. 68000 simultaneous measurements were found.

\begin{figure*}
\includegraphics[width=11cm]{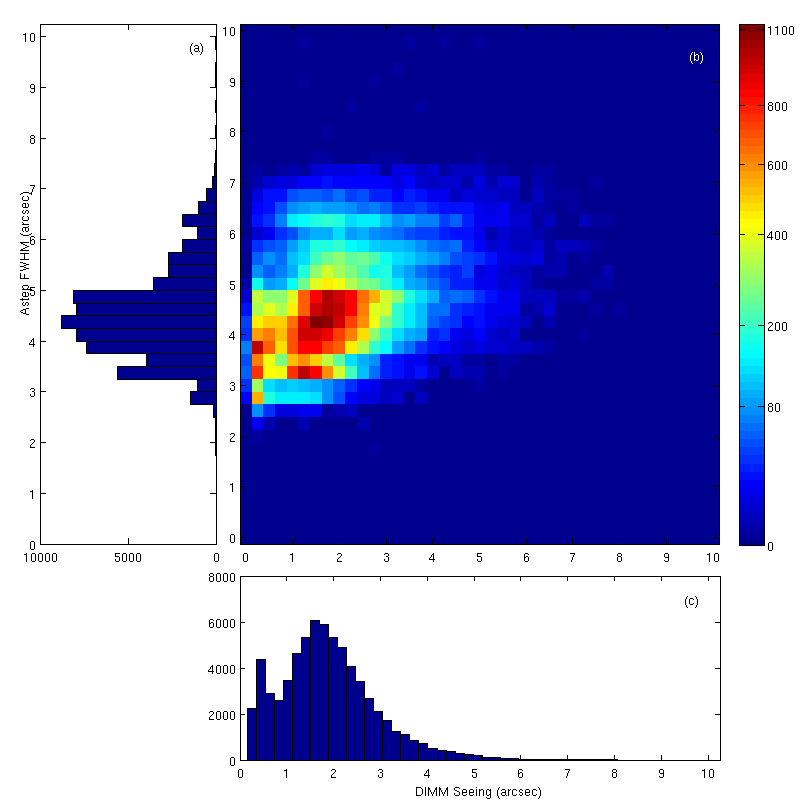}
\caption{ASTEP PSF size versus atmospheric seeing. (a) Histogram of the FWHM of ASTEP~400 images recorded during the winters 2010 and 2011. (c) Seeing as measured at the same time by the DIMM on the top of a 6m high platform. (b) Joint histogram of the seeing and the FWHM (colors correspond to the number of occurrences in the histogram).}
\label{fig:fwhm_seeing_joint_hist}
\end{figure*}

Figure~\ref{fig:fwhm_seeing_joint_hist} shows a histogram of co-occurrence of the seeing and the FWHM. Several observations can be made from this graph. (i) The FWHM is always greater than the seeing, which is expected.  (ii) For a given seeing, there is a large spread of the FWHM values, confirming that the seeing is not the only source of the PSF degradation. The correlation coefficient is 0.39.
(iii) The cloud of points exhibits a positive slope modelled by the quadratic fit
\begin{equation}
{\cal W}_{\rm ASTEP}=\left\{3.68^2 + \left(0.30 {\cal S}_{\rm DIMM}\right)^2\right\}^{1/2},
\label{eq:Wastep}
\end{equation}
with ${\cal W}_{\rm ASTEP}$ the ASTEP PSF in arcsec and ${\cal S}_{\rm DIMM}$ the DIMM seeing in arcsec. With separate measurements of DIMM telescopes on the ground and at 8m elevation, we estimate that the seeing at the level of ASTEP is generally $0.7"$ higher than measured at 8m. We thus use ${\cal S}={\cal S}_{\rm DIMM}-0.7"$. This relation is however valid only when the seeing is large enough, i.e. when the boundary layer is above both ASTEP and the DIMM (i.e. when ${\cal S}_{\rm DIMM}\wig{>}1"$). 
Using this eq.~\eqref{eq:Wastep} and a simple model for the seeing versus altitude derived from DIMM measurements at several elevations (Aristidi et al. 2009), we can predict a gain of approximately 0.2 arcsec on the median FWHM if we put ASTEP~400 at an elevation of 8\,m above the ground.

\subsection{Causes of PSF broadening}

We now turn to the analysis of the causes of the PSF broadening. We first calculate the Pearson correlation coefficient $r_{\rm Pearson}$ and the linear correlation between 
\begin{equation}
\Delta{\cal W}_1^2\equiv {\cal W}_{\rm ASTEP}^2-{\cal S}^2
\end{equation}
and various quantities $X^2$, where $X$ may be for example $T_{\rm M1}-T_{\rm out}$. We thus calculate by linear regression for each quantity $X$ values of $c$ and $a$ such that $\Delta{\cal W}_1^2\approx c^2+a^2X^2$. 

The results are indicated in table~\ref{tab:correlationsW1}. The factor $t$ is calculated as the ratio of the correlation coefficient $a^2$ and its variance. Among all the variables tested, the most significant correlation is with the temperature difference $T_{\rm M1}-T_{\rm out}$ for which we find a correlation coefficient $r=0.41$ and a slope of $0.25\,\rm arcsec/K$. Because perturbations to the PSF are expected to be additive and because the signal may be perturbed by other effects, we also select in each bin of 100 points the minimum of the highest 90 points, and perform a new regression analysis. We then obtain $c_{\Delta T_{M1}}=2.01$\,arcsec and $a_{\Delta T_{M1}}=0.23\,\rm arcsec/K$, a very similar result. 

\begin{table}
\caption{Correlation coefficients between ${\cal W}_1$ and various physical temperatures (see text). $a$ is in units of arcsec$/^\circ\rm C$.}
\label{tab:correlationsW1}
\begin{tabular}{lllll}
\hline\hline
Variable & $r_{\rm Pearson}$ & $t$ & $a^2$ & $a$ \\
\hline
$T_{\rm out}$&  -0.005&  -0.56&  -0.0001&    \\
$T_{\rm M1}-T_{\rm out}$&   0.410&  47.06&   0.0614&   0.2478 \\
$T_{\rm M2}-T_{\rm out}$&   0.333&  36.97&   0.3162&   0.5623 \\
$T_{\rm win}-T_{\rm out}$&   0.157&  16.59&   0.0052&   0.0718 \\
$T_{\rm box}-T_{\rm out}$&   0.173&  18.42&   0.0014&   0.0378 \\
$T_{\rm win}-T_{\rm box}$&   0.103&  10.79&   0.0019&   0.0441 \\
$T_{\rm fli}-T_{\rm box}$&   0.053&   5.50&   0.0078&   0.0884 \\
$T_{\rm cam}-T_{\rm box}$&   0.104&  10.95&   0.0198&   0.1406 \\
$T_{\rm ccd}-T_{\rm cam}$&   0.247&  26.64&   0.0031&   0.0552 \\
\hline
\end{tabular}
\end{table}

We now define an equivalent FWHM based on ${\cal W}_1$ but subtracted of the dependence in $T_{\rm M1}-T_{\rm out}$,
\begin{equation}
{\cal W}_2^2\equiv {\cal W}_1^2-c_{\Delta T_{M1}}^2-\left[a_{\Delta T_{M1}}\left(T_{\rm M1}-T_{\rm out}\right)\right]^2.
\end{equation}
We then perform the same regression analysis as previously but this time based on ${\cal W}_2$. The results are shown in table~\ref{tab:correlationsW2}. All the quantities that were significantly correlated with ${\cal W}_1$ now show a very weak correlation with ${\cal W}_2$, showing that the PSF was mostly due to turbulent fluctuations due to the temperature difference between the M1 mirror and the outside air.  

\begin{table}
\caption{Correlation coefficients between ${\cal W}_2$ and various physical temperatures (see text). $a$ is in units of arcsec$/^\circ\rm C$.}
\label{tab:correlationsW2}
\begin{tabular}{lllll}
\hline\hline
Variable & $r_{\rm Pearson}$ & $t$ & $a^2$ & $a$ \\
\hline
$T_{\rm out}$&  -0.037&  -3.91&  -0.0003&     \\
$T_{\rm M1}-T_{\rm out}$&   0.066&   6.96&   0.0085&   0.0923 \\
$T_{\rm M2}-T_{\rm out}$&   0.063&   6.59&   0.0511&   0.2261 \\
$T_{\rm win}-T_{\rm out}$&   0.068&   7.09&   0.0019&   0.0436 \\
$T_{\rm box}-T_{\rm out}$&   0.033&   3.46&   0.0002&   0.0153 \\
$T_{\rm win}-T_{\rm box}$&  -0.001&  -0.13&  -0.0000&    \\
$T_{\rm fli}-T_{\rm box}$&   0.033&   3.49&   0.0042&   0.0652 \\
$T_{\rm cam}-T_{\rm box}$&   0.028&   2.96&   0.0046&   0.0677 \\
$T_{\rm ccd}-T_{\rm cam}$&   0.086&   8.98&   0.0009&   0.0301 \\
$\delta T_{\rm M1}$&   0.099&  10.35&   0.0099&   0.0995 \\
\hline
\end{tabular}
\end{table}

In order to further test the influence of the out-of-focus observations we do the following calculation: We flag the moments when the telescope was re-focalised and measure the temperature evolution of the M1 mirror since that time and until a new focalisation as $\delta T_{\rm M1}$. We expect that when moving away from that temperature, the ASTEP FWHM should degrade. This is indeed confirmed by the correlation coefficients in table~\ref{tab:correlationsW2}, even though the correlation coefficient $r=0.1$ is relatively small. When considering the lower envelope of 90\% of the points, we find that the correlation coefficient increases to $r=0.43$ and with a constant $c_{\rm \delta T_{M1}}=1.96$\,arcsec and a slope $a_{\rm \delta T_{M1}}= 0.12\rm\,arcsec/K$. We thus define a new equivalent FWHM,
\begin{equation}
{\cal W}_3^2\equiv {\cal W}_2^2-c_{\delta T_{M1}}^2-\left[a_{\delta T_{M1}}\delta T_{M1}\right]^2.
\end{equation}

\begin{figure}
\includegraphics[width=\hsize]{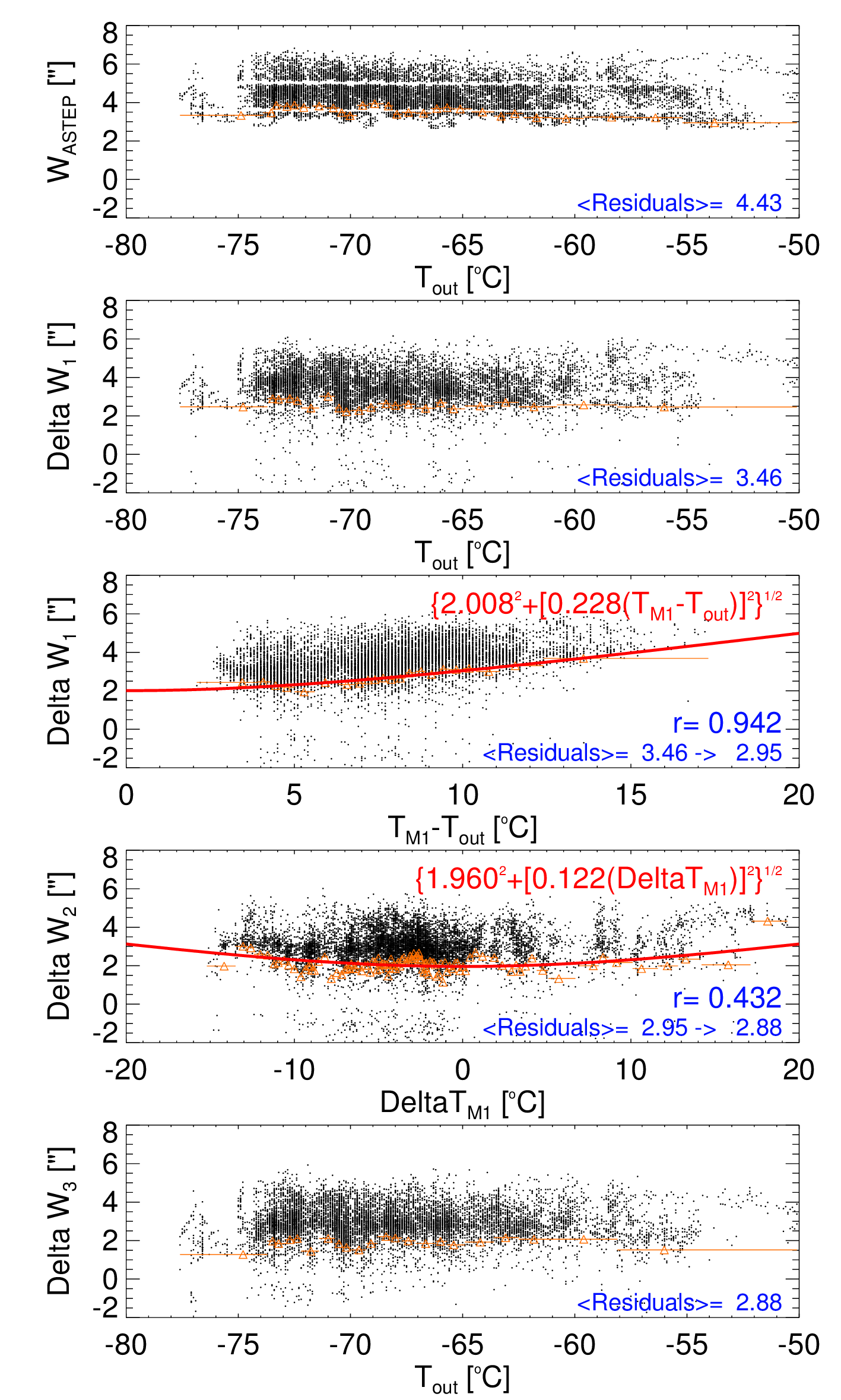}
\caption{Equivalent FWHM (in arcseconds) as a function of various temperatures (in Celsius) measured by ASTEP during 2010 and 2011. (See text for the definitions of ${\cal W}_1$, ${\cal W}_2$, ${\cal W}_3$ and $\delta T_{\rm M1}$.) The minimum envelope of 90\% of the points is indicated by orange triangles and corresponding horizontal lines (bin sizes). The red curves correspond to regression fits to this minimum envelope. The value of the Pearson correlation coefficient $r$ is indicated, as well as the mean residual (in arcsec). When a fit to the minimum envelope is calculated, mean residuals are indicated both for the data set and when subtracting the fit.}
\label{fig:wintercorrelations}
\end{figure}

Figure~\ref{fig:wintercorrelations} shows the different steps of the analysis. The top panel provides the measurements of ${\cal W}_{\rm ASTEP}$ which is mostly independent of outside temperature and has a mean value of $4.43$\,arcsec. Removing the inferred seeing ${\cal S}$ yields a mean residual of 3.46\,arcsec. The effect of M1 turbulence is obvious on the middle panel. Its removal yields ${\cal W}_2$ which has a mean of 2.95\,arcsec. Finally, the out-of-focus observations are responsible for a limited but still significant increase of the PSF size. The final equivalent FHWM ${\cal W}_3$ has a mean of 2.88\,arcsec.   

This value of $\langle{\cal W}_3\rangle$ is only slightly larger than the minimum value of the FWHM that we could obtain during the 2013 summer campaign. Although other effects are certainly present, we choose to stop the analysis at this point. We note that the slope that we have identified for the effect of the M1 seeing, i.e. $0.23\,\rm arcsec/K$ is very close to the value measured directly during the summer, i.e. $0.17\,\rm arcsec/K$. We were not able to identify a contribution due to the window seeing which may be hidden by the M1 turbulence signal. Finally, we were able to identify an effect due to the out-of-focus observations of about $0.12\,\rm arcsec/K$, in agreement with the value of $0.14\, \rm arcsec/K$ estimated from the summer campaign.

\section{Conclusions \& perspectives}
 
The low amount of precipitable water, excellent weather conditions and cold temperatures imply that the high plateaus of Antarctica are excellent sites for astronomy and in particular infrared astronomy. However, rapidly varying temperatures and a strong temperature gradient between the ground and higher atmospheric layers impose coping with temperature inhomogeneities on the optical path, dilatations of the instruments, and frost deposits. 

ASTEP~400 is a pilot telescope installed at the Concordia station, Dome C, Antarctica to perform precise photometric observations of large stellar fields in the visible and analyze its observations between years 2010 and 2013. The telescope ran smoothly since its installation in 2010 and could obtain continuous lightcurves with an excellent photometric precision, as illustrated by the detection of the secondary eclipse of WASP-19b \citep{Abe+2013}. However, the observations are characterized by a PSF which is larger than expected. We examined the reasons of this PSF broadening. 

A first reason is the installation of the telescope only about 2 meters above ground level, deep in the atmospheric surface boundary layer. This layer is characterized by a large turbulence level with a median seeing of 2\,arcsec. Using both our data acquired during the winter season and controlled experiments during the short summer season, we identified that the heating of the mirrors M1 and M2, necessary to prevent frost, is the main cause of the extra-broadening. The turbulence generated was shown to yield a PSF size proportional to the temperature difference between the mirror and the ambiant air equal to about 0.23\,$\rm arcsec/K$. This value is in good agreement with measurements obtained on mid-latitude telescopes. 

Another source of extra broadening is due to dilatations of the telescope, at a rate of about $0.14\,\rm arcsec/K$. The temperature difference between the mirrors and the ambiant air was between 5 and $10^\circ $C during the winter season implying a substantial overall broadening of the PSFs. Conversely, the very large temperature variations present at several locations on the optical path, i.e., at the entrance window of the camera box, between the upper and the lower camera box compartments, and at the entrance of the CCD camera had a smaller effect that could not be quantified. Similarly, jitter implied a small broadening with a standard deviation of only 0.3 arcsec in both directions. 

For future telescopes in Antarctica requiring small PSFs, these difficulties can be mitigated with the following approach: Mirrors should be ventilated, preferably with air extracted from ground level. The focal plane should be adjusted in real time. Tests performed with ASTEP~400 during summer 2013 were very promising in this respect. Furthermore, the installation of telescopes above the turbulent layer can strongly reduce the atmospheric seeing. For example, at Concordia, installing ASTEP at 8 meters elevation would reduce the seeing by $\sim\!0.7$\,arcsec. Installing it above 20-30 meters would further reduce the atmospheric seeing possibly down to a mere $0.5$\,arcsec. However, the instrument would have to cope with even stronger and faster temperature variations and more wind than experienced with ASTEP. 

Finally, we estimate that the installation of a tip-tilt system would be highly beneficial in reducing the contribution of the atmospheric seeing without requiring the telescope to be installed high above the ground. On the high Antarctica plateaus such as at Dome C, such a system would take advantage of the slow fluctuations of turbulence in the ground layer.

\acknowledgements
We are indebted to the constant logistical support of the {\em Institut Paul Emile Victor} and the {\em Programma Nazionale Di Ricerche in Antartide}, without which this project would not have been possible. ASTEP also benefited from the support of the {\em Agence Nationale de la Recherche}, {\em Institut National des Sciences de l'Univers} and of the {\em Programme National de Plan\'etologie}. We also wish to thank Luca Palchetti (INO-CNR Firenze, Italy) for kindly lending us an IR camera acquired through the PRANA and COMPASS projects.


\bibliography{astep_turbulence}
 


\appendix

\section*{Appendix: Basic considerations on thermal enclosures}\label{sec:appendix}


\section{A simple model}

The situation that we consider is simple. A system is kept at a warm temperature $T_{\rm int}$, while the outside temperature varies. A heating system (e.g. resistances) provides the heat flux necessary to keep the inside temperature at a relatively stable value. Several questions arise: what is the value of this heat flux? What are the remaining temperature gradients in the system, and in particular at the inner and outer surfaces of the system (it is from these that turbulence will arise due to convective plumes)? How does the system evolve with time?

In order to partially answer these questions, we develop a very simple model based on heat transport by radiation, conduction and convection. 

\subsection{Radiation}

Radiative transport is possible only in a transparent environment. The flux transported, averaged over all wavelength and assuming perfect transparency is
\begin{equation}
F_{\rm rad}=\sigma T^4,
\end{equation} 
where $\sigma$ is Stefan-Boltzmann's constant, and T is the temperature of the emitted radiation for a perfect blackbody. 
An opaque surface will generally cool radiatively at rate that is smaller by its emissivity $\epsilon$,
\begin{equation}
F_{\rm rad}=\epsilon\sigma T^4.
\end{equation} 
If irradiated, it will reflect a fraction $\sim\! (1-\epsilon)$ of the flux that it receives while absorbing $\sim\! \epsilon$ of that flux (assuming the layer is completely opaque to the radiation considered).

\subsection{Conduction}
Conductive transport is relatively straightforward. It depends on the material considered through its thermal conductivity $K$, generally expressed in $\rm W\,m^{-1}\,K^{-1}$. The heat flux conducted by any material is proportional to the temperature difference between its extremities,
 \begin{equation}
F_{\rm cond}=K{\Delta T\over \ell},
\end{equation}
where $\ell$  is the length over which the temperature difference $\Delta T$ holds.

\subsection{Convection}
Convection arises only if the temperature difference exceeds a certain value, that is generally proportional to the viscosity of the element considered. For our purposes we only need to consider the case of convection in gases. It is extremely difficult to estimate precisely how much heat will be transported by convection as it depends on many different factors among which there are the properties of the surfaces and fluids considered, the aspect ratio, the hydrodynamic background,...etc. 

In the case of free air above a heated surface, an empirical solution is to use a conducto-convective coefficient (h) and write
\begin{equation}
F_{\rm free conv}=h\Delta T.
\end{equation}
Typical values of h are of order $5\,\rm W\,m^{-2}\,K^{-1}$ for still air, up to $20\,\rm W\,m^{-2}\,K^{-1}$  for air flowing above a surface (REFERENCE). Note that a comparison to the thermal conductivity of air $K\sim 0.02 \rm W\,m^{-1}\,K^{-1}$, this is equivalent to a boundary layer of thickness of 4 to 1 mm. 

The case of convection between two glass layers with different temperatures corresponds to the well-studied Rayleigh-B\'enard convection. The convection will appear if the Rayleigh number,
\begin{equation}
Ra\equiv {g \beta (T_2 - T_1)\over \nu\kappa}d^3,
\end{equation}
becomes larger than a critical value $Ra_{\rm crit}$. $g$ is the gravity, $d$ the distance between the two layers, $\beta$ the thermal dilatation coefficient ($\beta=1/T$ for a perfect gas), $\nu$ is the kinematic viscosity, and $\kappa$ the thermal diffusivity ($\kappa=K/(\rho c_p)$).  The critical Rayleigh number to achieve convection is larger for boxes with a small aspect ratio (e.g., $\sim\!4000$ for a 1:1 box), and smaller for infinitely long plates (around $600$ in the theoretical limit). For our purposes, we will use $Ra_{\rm crit}\sim 1700$. 

Before reaching the critical Rayleigh number, heat is transported by conduction. Given a fixed temperature difference, the heat flux is inversely proportional to the thickness of the layer. When convection sets in, we assume that the Nusselt number that measures the ratio between the total transported heat flux to the conductive heat flux is 
\begin{equation}
Nu=(Ra/Ra_{\rm crit})^{1/3}.
\end{equation}
The flux transported is thus
\begin{equation}
F=\left(Ra\over Ra_{\rm crit}\right)^{1/3} {K\over d}\Delta T,
\end{equation}
or
\begin{equation}
F=\left(g\beta K^2 \Delta T^4\over \nu Ra_{\rm crit}\right)^{1/3},
\end{equation}
and becomes hence independent of the thickness of the convecting layer $d$. 

We can also turn the problem around and ask when does the layer start to become convective? 
This occurs when $Ra=Ra_{\rm crit}$, i.e. when
 \begin{equation}
d=\left({\nu K\over g\beta}{Ra_{\rm crit}\over \Delta T}\right)^{1/3}
\end{equation}
With typical values for dry air, 250K and $\Delta T=10$\,K, we obtain d=2.9mm. 
Increasing the separation beyond that distance thus does not increase the insulating properties of the material.

\section{Simple simulations}

The solution of the problem is analytical as long as one neglects or linearizes the radiative transfer part. But even in this case it becomes rapidely quite complicated when one increases the number of layers (e.g. for double glass). We therefore use a simple numerical solution to the problem that also accounts for fluctuations in the external temperature profile. 

In order to simulate the instrument in a simple way, we consider a small number of layers that are characterized by their bottom, top and central temperature. We assume that the temperature profile varies linearly inside any given layer (so that the central temperature is really the median between bottom and top temperatures). We assume that the temperature profile is continuous.

The code first calculates the fluxes across each layer as a function of their bottom and top temperatures. This flux accounts for conduction, radiation and convection, when necessary (using the Rayleigh criterion described above).

It then solves implicitely the new temperature of each layer after a time step $\Delta t$, such that
\begin{equation}
T(t)-T(t-\Delta t)={\Delta F(t)\over c_P d}\Delta t,
\end{equation}
where $\Delta F(t)$ is the net flux across the layer (accounting for possible heating of the bottom and/or top layers, and the loss of heat in the layer considered). 
(Note that a fully implicit scheme is not necessarily the best numerical alternative, but it was chosen for its simplicity). 

This net flux is calculated as
\[
\Delta F=F_{i-1}-F_i,
\]
where i corresponds to the layer considered. This is equivalent to an upwind scheme but using the assumption that transport is dominantly outwards. (Clearly, a refined model solving the full heat transfer problem with some spatial resolution would be desirable, yet our simplified model should be sufficient to provide orders of magnitude estimates, given that it is in any case accurate in the static case). 

In a solid layer (e.g. glass), $F$ is simply calculated using the equation for conduction as a function of the temperature of the top and bottom of the layer. In a gaseous layer (e.g. air, vacuum), $F$ is a sum of a conductive/convective flux and of a radiative flux. This net radiative flux is a function of the emissivities $\epsilon$, and temperatures $T$ of the solid layers directly below and above the one considered, respectively. 

Considering n reflections, one can show that the net radiative flux writes (using $R=1-\epsilon$)
\begin{eqnarray}
 F_{\rm rad} &\!=\!&\epsilon_1 \sigma T_1^4 (1-R_2)(1+R_1 R_2 +...+R_1^n R_2^n) \nonumber\\
&&- \epsilon_2 \sigma T_2^4 (1-R_1)(1+R_1 R_2 +...+R_1^n R_2^n). 
\end{eqnarray}
In what follows, we consider 4 reflections.

\begin{figure}
\includegraphics[width=\hsize]{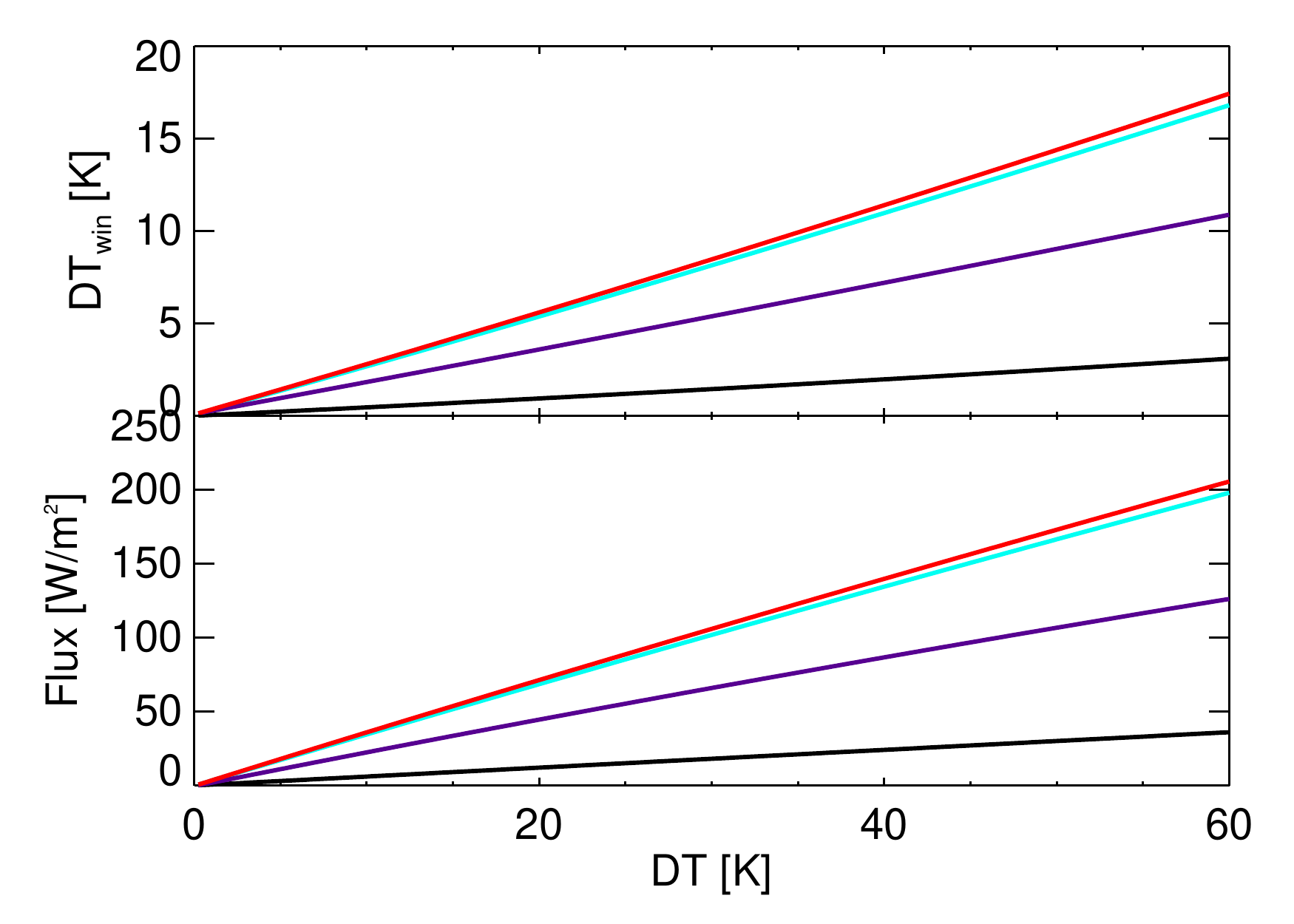}
\caption{Insulation efficiency of different materials as a function of $\Delta T$ the difference between the outside (ambient) temperature and inside (e.g. camera box) temperature. {\em Top panel}: Value of the temperature difference between the outside of the material and the ambient air. {\em Bottom panel}: Heat power required to maintain the inside temperature at $-20\celsius$. The different curves correspond to 4\,cm of roofmate/polystyrene (black), 1\,cm of titanium (red), 1\,cm of BK7 glass (blue) and a double glass window made two layers of 1\,cm-thick BK7 glass separated by 5\,mm of air (purple).}
\label{fig:thermalmulti}
\end{figure}

Figure~\ref{fig:thermalmulti} shows the insulation efficiency of different materials, in terms of the temperature difference between the outside surface and the ambient air, and in terms of the heat flux necessary to maintain a given temperature (chosen to be equal to $-20\celsius$ in agreement with the nominal setup for the ASTEP~400 camera box). Unsurprisingly, glass and metal (here titanium) are relatively bad insulators, while roofmate/polystyrene is an efficient one. Using a double glass window allows reducing the temperature jump on the outside of the window -and this was hence the solution chosen for ASTEP~400. (Making a double glass window with argon or vacuum would have been problematic in the Antarctic environment and was not retained in spite of it being more efficient). 

Given the maximum $\sim\!120\,$W electrical consumption of the ASTEP science camera (most of it presumably being turned into heat) and the relatively small area of the box ($\sim\! 0.9\rm\,m^2$), no extra heating is required even in the coldest days of the winter. However, the temperature fluctuations along the optical path can generate turbulence and thus affect the quality of the images being taken.

\section{Modification of the curvature radius of lenses and mirrors}

Let us now consider the possibility that a lens will be deformed due to changes of the temperatures. For simplicity, we will assume that the lens has parallel faces, of length $l_1$ and $l_2$, respectively, and a thickness $d$. Its radius of curvature $R$ is measured from $l_1$, and we assume that any curvature is spherical. If we consider the angle $\alpha$ from which the lens is seen at its focal point, then it is easy to show that 
\begin{equation}
\begin{cases}
l_1=\alpha R, & \\
l_2=\alpha(R+d). & 
\end{cases}
\end{equation}
This implies
\begin{equation}
R=d/(l_2/l_1-1).
\end{equation}
Let us consider that all lengths will be affected by a temperature change $\Delta T$ according to $\Delta l=\chi \Delta T$, where $\chi$ is the coefficient of thermal expansion (CTE) of the lens material. We are mostly interested in changes in the curvature radius $R$ as a function of the temperature change, and also as a function of whether the change is homogeneous or heterogeneous. 


\subsection{Consequence of a homogeneous temperature change}

A temperature change that is homogeneous in the lens changes all its lengths by $\chi\Delta T$. In the plane-parallel case, this has no consequence. In the spherically-curved lens case, this implies that the new curvature radius $R$ changes compared to the old one $R_0$ proportionally to the change in the length of the lens thickness $d$,
\begin{equation}
\Delta R/R=\chi \Delta T.
\end{equation}
For example, this implies that between the building of ASTEP~400 at $15\celsius$ and its operation at $-65\celsius$ the curvature radius change, for a BK7 glass with $R_0$=15\,cm will amount to $\sim\! 91\rm\,\mu m$. 

\subsection{Consequence of a heterogeneous vertical temperature variation}

Let us now consider that the temperature of face 1 is different from that of face 2. For simplicity, we will consider that only face 2 (e.g. outside) is affected, by a temperature change $\Delta T_{12}$. In this case, a plane-parallel lens will curve so that both faces remain parallel with the same (non-infinite) curvature radius. This new curvature radius is
\begin{equation}
R+\Delta R={d\over l_2 (1+\chi \Delta T_{12})/l_1-1},
\end{equation}
or
\begin{equation}
R+\Delta R={R\over 1+\chi \Delta T_{12}(1+R/d)}.
\end{equation}
In the limit of small dilatations ($\chi  \Delta T_{12} \ll 1$), one can easily show that
\begin{equation}
{\Delta R\over R}\approx -\chi \Delta T_{12} (1+R/d).
\end{equation}
It is important to note that the change in focal radius is now proportional to R/d which is often a factor that is much larger than unity (10 to 100, typically). 

For ASTEP~400, the change in R is likely to be more problematic. With BK7 glass, and $d=1.5\,$cm, $R\sim 20$\,cm, $\chi=7.6\times 10^{-6}\rm\,K^{-1}$, $\Delta T_{12}=3$\,K, the above relation yields a change of the focal length of 50$\,\mu$m. Given the temperature variations experienced in the glass as a result of outside temperature changes (see previous sections), there is at least an unavoidable variable 1\,K temperature change that translates into a 16.7$\,\mu$m change of the focal length.

\subsection{Consequence of a heterogeneous horizontal temperature variation}

We now consider that a ring used to maintain the two lenses precisely at the required distance will introduce a horizontal heterogeneity in the temperature of the lenses, and hence variations of the curvature. Contrary to the previous cases, this cannot be corrected by an autofocus because it affects only part of the image.
Very roughly, this horizontal temperature change should be at most $\Delta T_{\rm h}=5$\,K. As in the homogeneous temperature variation case, one should be able to estimate variations in the curvature as
\begin{equation}
{\Delta R\over R}\approx -\chi \Delta T_{\rm h},
\end{equation}
i.e., absolute variations of 8 microns for a curvature of 20cm. Given that ASTEP has a camera with pixels of 9 microns, horizontal temperature variations of lenses are not expected to affect the observations.

\end{document}